\documentclass[aps,prb,reprint,superscriptaddress,nobibnotes]{revtex4-1}
\usepackage{graphicx,epsfig}
\usepackage{amsmath,amssymb,bm}
\usepackage{color}
\usepackage[usenames,dvipsnames]{xcolor}
\usepackage[colorlinks=true,linkcolor=Maroon,citecolor=OliveGreen,urlcolor=Blue,linktoc=page]{hyperref}
\usepackage{placeins}
\usepackage{ulem}
\usepackage[symbol]{footmisc} %
\usepackage{multirow}

\begin{document}


\title{Nature of the ferromagnetic-antiferromagnetic transition in Y$_{1-x}$La$_{x}$TiO$_{3}$}    
\author{S. Hameed}
\thanks{Corresponding authors: hamee007@umn.edu, greven@umn.edu}
\affiliation{School of Physics and Astronomy, University of Minnesota, Minneapolis, MN 55455, U.S.A.}
\author{S. El-Khatib}
\affiliation{Department of Chemical Engineering and Materials Science, University of Minnesota, Minneapolis, MN 55455, U.S.A.}
\affiliation{Department of Physics, American University of Sharjah, P.O. Box 26666, Sharjah, United Arab Emirates}
\author{K. P. Olson}
\affiliation{Department of Chemical Engineering and Materials Science, University of Minnesota, Minneapolis, MN 55455, U.S.A.}
\author{B. Yu}
\affiliation{School of Physics and Astronomy, University of Minnesota, Minneapolis, MN 55455, U.S.A.}
\author{T. J. Williams}
\affiliation{Neutron Scattering Division, Oak Ridge National Laboratory, Oak Ridge, Tennessee 37831, USA}
\author{T. Hong}
\affiliation{Neutron Scattering Division, Oak Ridge National Laboratory, Oak Ridge, Tennessee 37831, USA}
\author{Q. Sheng}
\affiliation{Department of Physics, Columbia University, New York, New York 10027, USA.}
\author{K. Yamakawa}
\affiliation{Department of Physics, Columbia University, New York, New York 10027, USA.}
\author{J. Zang}
\affiliation{Department of Physics, Columbia University, New York, New York 10027, USA.}
\author{Y. J. Uemura}
\affiliation{Department of Physics, Columbia University, New York, New York 10027, USA.}
\author{G. Q. Zhao}
\affiliation{Department of Physics, Columbia University, New York, New York 10027, USA.}
\affiliation{Institute of Physics, Chinese Academy of Sciences, Beijing 100190, China.}
\author{C. Q. Jin}
\affiliation{Institute of Physics, Chinese Academy of Sciences, Beijing 100190, China.}
\author{L. Fu}
\affiliation{Department of Physics, Zhejiang University, Hangzhou 310027, China.}
\author{Y. Gu}
\affiliation{Department of Physics, Zhejiang University, Hangzhou 310027, China.}
\author{F. Ning}
\affiliation{Department of Physics, Zhejiang University, Hangzhou 310027, China.}
\author{Y. Cai}
\affiliation{Department of Physics and Astronomy, University of British Columbia, Vancouver, British Columbia V6T 1Z1, Canada}
\affiliation{Stewart Blusson Quantum Matter Institute, University of British Columbia, Vancouver, British Columbia V6T 1Z4, Canada}
\affiliation{TRIUMF, Vancouver, British Columbia, V6T 2A3, Canada.}
\author{K. M. Kojima }
\affiliation{TRIUMF, Vancouver, British Columbia, V6T 2A3, Canada.}
\author{J. W. Freeland}
\affiliation{X-ray Science Division, Argonne National Laboratory, Argonne, IL 60439, USA}
\author{M. Matsuda}
\affiliation{Neutron Scattering Division, Oak Ridge National Laboratory, Oak Ridge, Tennessee 37831, USA}
\author{C. Leighton}
\affiliation{Department of Chemical Engineering and Materials Science, University of Minnesota, Minneapolis, MN 55455, U.S.A.}
\author{M. Greven}
\thanks{Corresponding authors: hamee007@umn.edu, greven@umn.edu}
\affiliation{School of Physics and Astronomy, University of Minnesota, Minneapolis, MN 55455, U.S.A.}


\widetext
\date{\today}

\begin{abstract}
We explore the magnetically-ordered ground state of the isovalently-substituted Mott-insulator Y$_{1-x}$La$_{x}$TiO$_{3}$ for $x$ $\leq$ 0.3 via single crystal growth, magnetometry, neutron diffraction, x-ray magnetic circular dichroism (XMCD), muon spin rotation ($\mu$SR) and small-angle neutron scattering (SANS). 
We find that the decrease in the magnetic transition temperature on approaching the ferromagnetic (FM) - antiferromagnetic (AFM) phase boundary at the La concentration $x_c$ $\approx$ 0.3 is accompanied by a strong suppression of both bulk and local ordered magnetic moments, along with a volume-wise separation into magnetically-ordered and paramagnetic regions. The thermal phase transition does not show conventional second-order behavior, since neither a clear signature of dynamic critical behavior nor a power-law divergence of the magnetic correlation length is found for the studied substitution range; this finding becomes increasingly obvious with substitution.
Finally, from SANS and magnetometry measurements, we discern a crossover from easy-axis to easy-plane magneto-crystalline anisotropy with increasing La substitution. These results indicate complex changes in magnetic structure upon approaching the phase boundary.
\end{abstract}
\pacs{}
\maketitle


\section{Introduction}
Perovskite and perovskite-derived transition metal oxides provide excellent platforms to study quantum many-body phenomena \cite{Imada1998}. 
In their undoped parent state, these materials are often Mott-Hubbard or charge-transfer insulators as a result of strong electronic correlations \cite{Zaanen1985,Torrance1991}. Arguably the most prominent examples are the lamellar cuprates, which in their undoped state are AFM charge-transfer insulators with square-planar CuO$_2$ sheets and Cu 3$d^9$ spin-$\frac{1}{2}$ degrees of freedom \cite{Keimer2015}. Another prominent example are the cubic rare-earth (R) titanates RTiO$_3$, which are Mott insulators with a spin-$\frac{1}{2}$ $3d^{1}$ electronic configuration in their undoped state and realize both AFM and FM ground states \cite{Mochizuki2004}. Similar to the cuprates, an insulator-metal transition can be induced in RTiO$_3$ $via$ charge-carrier doping \cite{Mochizuki2004}. Unlike the cuprates, the rare-earth titanates are not known to exhibit a superconducting phase, and their orbital degrees of freedom are unquenched. As a result of a strong coupling of orbital  and spin degrees of freedom, the RTiO$_3$ compounds exhibit a variety of spin-orbital structures with complex electronic and magnetic properties and phases \cite{Mochizuki2004}. 

RTiO$_{3}$ compounds exhibit a GdFeO$_{3}$-type distorted perovskite structure that involves tilts and rotations of the TiO$_{6}$ octahedra, with a resultant Ti-O-Ti bond angle of less than $180^{\rm o}$. A decrease in R-site ionic radius increases this distortion and thereby decreases the Ti-O-Ti bond angle \cite{MacLean1979}, which controls the coupling between neighboring Ti ions and drives the system through an AFM to FM transition \cite{Greedan1985,Zhou2005a}. 
While continuous control of the R ion radius (and hence of the Ti-O-Ti bond angle) is not possible due to the discrete choices for the R ion, a similar transition is known to occur in solid-solutions such as Sm$_{1-x}$Gd$_{x}$TiO$_{3}$ \cite{Amow2000} and Y$_{1-x}$La$_{x}$TiO$_{3}$ \cite{Goral1982,Okimoto1995,Zhou2005}. Importantly, the latter system is relatively simple, as the rare-earth ions are non-magnetic, and hence the magnetic properties arise solely from the Ti$^{3+}$ ion. Moreover, the low neutron absorption cross-sections of Y and La (unlike those of Sm and Gd) enable a straightforward study of the magnetic ground state evolution with substitution via neutron diffraction. Powder x-ray diffraction of Y$_{1-x}$La$_{x}$TiO$_{3}$ reveals an increase in cell volume with increasing $x$, reflecting an increasing Ti-O-Ti bond angle \cite{Goral1982}. Magnetic susceptibility measurements show a sharp suppression of the magnetic phase transition temperatures ($T_{C}$ and $T_{N}$) near $x_{c}$ $\approx0.3$ \cite{Goral1982,Okimoto1995,Zhou2005}. A theoretical study predicts a quantum critical point at this substitution level \cite{Khaliullin2002}. Apparent experimental support for this possibility comes from specific heat, magnetic susceptibility (under hydrostatic pressure), and dielectric constant results for single crystals (for $x \leq 0.3$), which were interpreted as indicative of a spin-orbital-liquid state between the FM and AFM phases \cite{Zhao2015}. More recent work focussed on the lattice and magnetic dynamics in polycrystalline samples of Y$_{1-x}$La$_{x}$TiO$_{3}$ \cite{Li2016}. However, the nature of the evolution of the magnetic ground state is still under debate, and most studies, particularly neutron scattering measurements of single crystals, have been largely restricted to the end compounds YTiO$_{3}$ \cite{Ulrich2002} and LaTiO$_{3}$ \cite{Lynn2002}.

In this paper, we re-examine the evolution with $x$ of the magnetic properties of Y$_{1-x}$La$_{x}$TiO$_{3}$. Single-crystal samples are comprehensively studied via neutron and x-ray scattering, magnetometry, and $\mu$SR. Along with a strong suppression of the Curie temperature on approaching $x_{c}$ $\approx0.3$, we find evidence for a suppression of the local and bulk ordered magnetic moments, accompanied by phase separation into magnetically-ordered and paramagnetic regions. Furthermore, we observe no evidence for dynamic critical behavior on the $\mu$SR timescale ($\sim$ 1~$\mu$s), which indicates that the thermal phase transition is not conventional second-order. This conclusion is further substantiated by the absence of a power-law divergence of the magnetic correlation length, as evidenced by SANS measurements. The weakly first-order nature of the transition becomes increasingly obvious with increasing La substitution. Finally, magnetometry and SANS experiments reveal that the system crosses over from easy-axis to easy-plane magneto-crystalline anisotropy on approaching the phase boundary.

This paper is organized as follows: Section~\ref{section:expmeth} describes the single-crystal growth and characterization, as well as experimental details. Sections~\ref{section:magnetometry} - \ref{section:comparison} describe measurements of the average bulk ordered magnetic moment and compare the results obtained with different probes. Section~\ref{section:MuSR} details $\mu$SR measurements of the magnetic volume fraction, local ordered moment and spin-lattice relaxation rate, and compares these results to those obtained with the bulk probes in Sections~\ref{section:magnetometry} - \ref{section:comparison}. Section~\ref{section:muSRvsNeutron} compares the $\mu$SR and neutron diffraction results with a specific focus on the nature of the thermal phase transition. Section~\ref{section:sans} discusses the SANS measurements and their relation to the neutron diffraction and $\mu$SR results. 
Section~\ref{section:easyaxis} describes the investigation of magnetocrystalline anisotropy and its implications for the bulk- and local-moment results in Sections~\ref{section:magnetometry} - \ref{section:MuSR}. Finally, we summarize our main conclusions in Section~\ref{section:concl}.


\section{Experimental methods}
\label{section:expmeth}

Single crystals of Y$_{1-x}$La$_{x}$TiO$_{3}$ with $x$ $\leq$ 0.3 were melt-grown with the optical floating-zone technique. The starting materials were La$_{2}$O$_{3}$, Y$_{2}$O$_{3}$ and Ti$_{2}$O$_{3}$. Due to the instability of Ti$^{3+}$ in Y$_{1-x}$La$_{x}$TiO$_{3}$, which tends to oxidize to Ti$^{4+}$ in the presence of traces of oxygen, an oxygen-deficient \textit{starting} composition Y$_{1-x}$La$_{x}$TiO$_{3-\delta}$ was used. This was achieved by adding Ti powder to the starting materials. The La$_{2}$O$_{3}$ and Y$_{2}$O$_{3}$ powders were pre-dried in air at 1000$^{\text{o}}$C for 12 hours. The starting materials were then mixed in the stoichiometric ratio Y$_{1-x}$La$_{x}$TiO$_{3-\delta}$, and subsequently pressed at 70 MPa into two rods with a diameter of 6 mm and lengths of 20 mm and 100 mm, for use as seed and feed rods, respectively. The feed and seed rods were then loaded into a Crystal Systems, Inc. four-mirror optical image furnace. The growths were performed at a rate of 5 mm/h, in a reducing atmosphere with a mixed gas of 5\%H$_{2}$/95\%Ar at a pressure of 5 bar. For $x = 0$ and $x = 0.1$, several $\delta$ values in the range 0 - 0.08 were attempted. Since it is known that oxygen off-stoichiometry tends to decrease $T_{C}$ \cite{Ulrich2002}, the final value of $\delta$ was chosen so as to maximize $T_{C}$. It was found that $\delta$ = 0.04 gave the highest (optimal) $T_{C}$ for both $x = 0$ and $x = 0.1$; $\delta$ was fixed at this value thereafter at higher La concentrations. Attempts to grow single crystals above $x = 0.3$ failed and yielded only polycrystalline material, even though several values of $\delta$ were attempted. The grown samples were characterized with Laue x-ray diffraction to confirm single crystallinity. The chemical compositions of the samples, determined using wavelength-dispersive spectroscopy (WDS), are displayed in Table~\ref{tab:table1} and seen to be in good agreement with the nominal compositions. Note that the oxygen off-stoichiometry cannot be obtained from WDS and was not determined for our samples.

\begin{table}[h!]
  \begin{center}
    \caption{Nominal and measured compositions of the Y$_{1-x}$La$_{x}$TiO$_{3}$ single crystals, and the saturation magnetic moments at 6 K estimated from the exponential extrapolation of the SQUID magnetometry results in Fig.~\ref{fig:Magn} (c).}
    \label{tab:table1}
    \vspace*{5mm}

    \begin{tabular}{c|c|c|c|c}
       \textbf{Nominal} & \textbf{Actual} & \textbf{Nominal} &  \textbf{Measured} & \textbf{FM} \\
       \textbf{x} & \textbf{x} & \textbf{(Y+La):Ti} &  \textbf{(Y+La):Ti} & \textbf{moment} \\
         &   & \textbf{ratio} &  \textbf{ratio} & ($\mu$$_{\text{B}}$/Ti) \\

      \hline
      0 & - & 1:1 & 0.968(13):1 & 0.97(2)\\
      0.10 & 0.107(11) & 1:1 & 0.984(14):1 & 0.63(1)\\
      0.15 & 0.153(9) & 1:1 & 0.983(14):1 & 0.55(1)\\
      0.20 & 0.213(23) & 1:1 & 0.988(35):1 & 0.40(1)\\
      0.25 & 0.264(9) & 1:1 & 0.971(15):1 & 0.34(1)\\
      0.30 & 0.312(24) & 1:1 & 0.988(39):1 & 0.29(1)
      
    \end{tabular}
  \end{center}
\end{table}

\begin{figure*}
\includegraphics[width=0.9\textwidth]{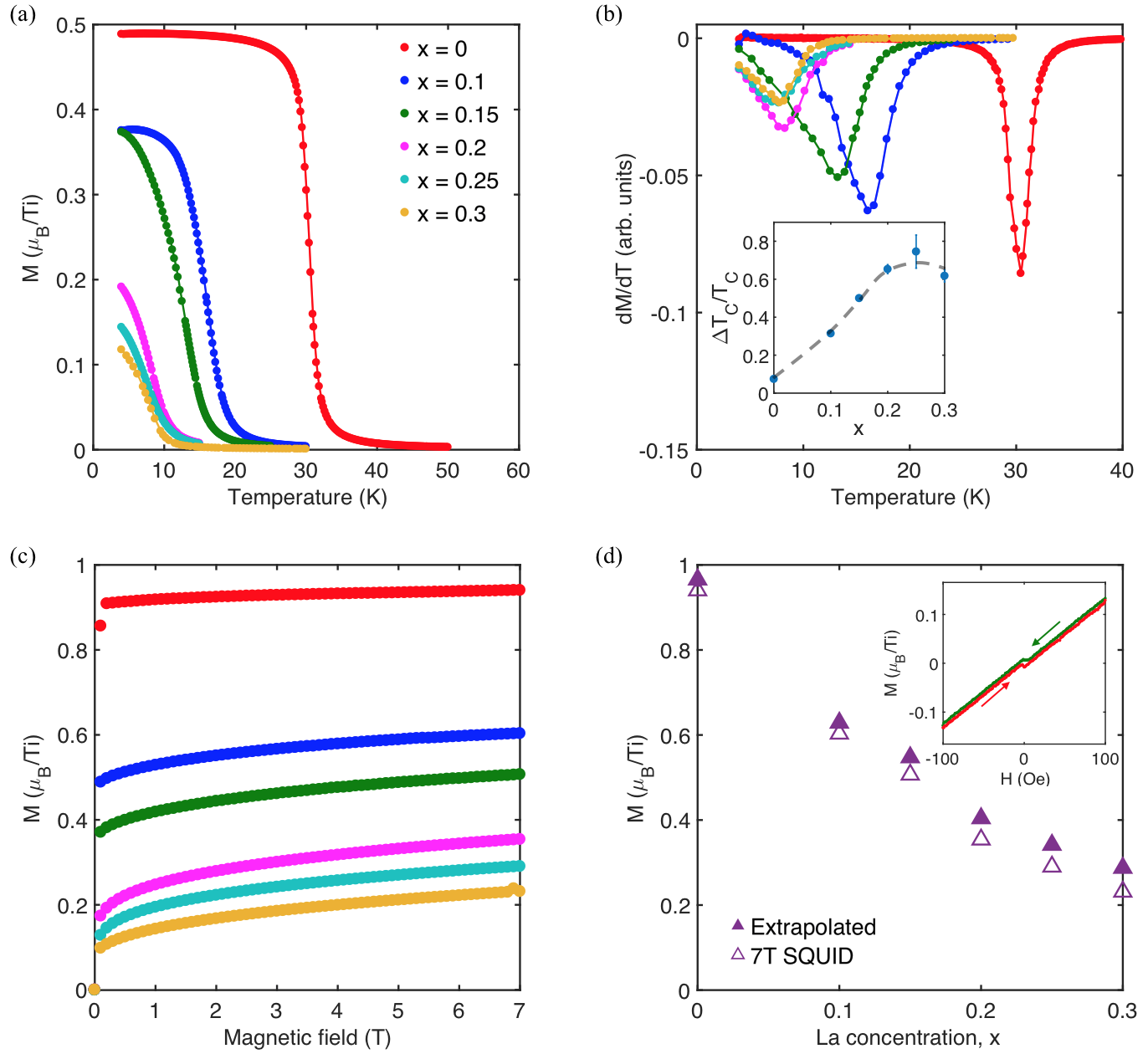}
\caption{Temperature and La concentration dependence of (a) the field-cooled magnetization $M$ and (b) $dM/dT$, obtained with an applied $c$-axis magnetic field of 500 Oe, for Y$_{1-x}$La$_{x}$TiO$_{3}$ ($ 0 \leq x \leq 0.3$) single crystals . The lines are guides to the eye. The inset in (b) shows the ratio of the full-width at half maximum ($\Delta T_C$) of the peaks in (b) to the $T_C$ value extracted from the peak position. The dashed line is a guide to the eye. (c) Magnetic-field dependence of the magnetic moment per Ti ion along the $c$-axis at $T = 6$ K. The samples were zero-field cooled and measured in ascending field. (d) Comparison of the measured magnetic moments at 7 T with the saturation magnetic moments estimated from an exponential extrapolation of the data above 2 T in (c). The inset shows the hysteresis loop for YTiO$_{3}$ at 5 K. Note that in the forward and reverse sweeps, the field was swept to 0.5 T and -0.5 T respectively, before reversing the sweep direction. However, only a portion of the data is shown in order to highlight the zero-remanence in the system.}
\label{fig:Magn}
\end{figure*}

Magnetic-susceptibility measurements were performed with a Quantum Design, Inc. dc SQUID magnetometer. The samples were pre-aligned with a Laue diffractometer and the magnetic field was applied along the $c$-axis. 

Neutron diffraction experiments were performed on the HB-1 thermal triple-axis spectrometer at the High Flux Isotope Reactor, Oak Ridge National Laboratory. The samples were single crystals with volumes of approximately $0.2 - 0.4$ cm$^3$. Some of the samples were measured at the CG-4C cold triple-axis spectrometer. One of the samples was measured at both HB-1 and CG-4C and the data were used to obtain the scale factor required to compare data acquired at the two different instruments. A $^{4}$He-cryostat was used to reach temperatures down to 1.5 K. The fixed final energy of the scattered neutrons and the collimations were 13.5 meV and $48'-80'-80'-120'$ at HB-1, and 4.5 meV and $\text{open}-\text{open}-80'-\text{open}$ at CG-4C. All data were obtained in the ($0KL$) scattering plane. PG filters and a cooled Be filter were used at HB-1 and CG-4C, respectively, to eliminate higher-order neutrons.

Small-angle neutron scattering data were obtained on the NG7 SANS instrument at the NIST Center for Neutron Research, using cold neutrons with a wavelength of 5.2 \AA~(neutron energy of 3.03 meV). 
The data were taken with two sample-detector distances, 2.5 m and 15.8 m, in order to extend the range of accessible scattering vectors (0.002 \AA$^{-1}$ $<q<$ 0.2 \AA$^{-1}$).
X-ray absorption spectroscopy/x-ray magnetic circular dichroism (XAS/XMCD) measurements using total fluorescence yield (TFY) detection mode were carried out at beam-line 4-ID-C of the Advanced Photon Source, Argonne National Laboratory. The sample surfaces were cut parallel to the (001) plane and polished to a surface roughness of $\sim$ 0.3 $\mu$m using polycrystalline diamond suspension.

The $\mu$SR measurements were performed on the M20 surface muon beam line using the LAMPF spectrometer at the Centre for Molecular and Materials Science at TRIUMF in Vancouver, Canada. The initial muon spin was anti-parallel to its momentum, and the sample [001] axis was orientated approximately perpendicular to the muon spin direction. A helium gas-flow cryostat was used to control the temperature down to 2 K.


\section{Results}
\label{section:bulklocal}

The parent material YTiO$_{3}$ has a non-trivial magnetic structure with predominant FM-aligned spins along the $c$-axis, and additional G-AFM and A-AFM components along the $a$- and $b$-axes, respectively, due to spin-canting \cite{Ulrich2002}. Here, we investigate the substitution dependence of the magnetic ground state of Y$_{1-x}$La$_{x}$TiO$_{3}$ using five complementary techniques: magnetometry, XAS/XMCD, triple-axis neutron diffraction, $\mu$SR and SANS.


\subsection{Magnetometry}
\label{section:magnetometry}

Figure~\ref{fig:Magn} (a,b) shows the substitution dependence of the field-cooled magnetization $M$ and its temperature derivative $dM/dT$ in an applied magnetic field of 500 Oe. A strong suppression of the Curie temperature, $T_{C}$, defined here as the peak position of $dM/dT$, is observed with increasing La substitution. Whereas YTiO$_{3}$ exhibits $T_{C}$ $\sim$ 30 K, the transition temperature decreases to $\sim$ 16 K for $x = 0.1$ and $\sim$ 12 K for $x = 0.15$, and then levels off  at $\sim$ 6 - 7 K for $x = 0.2 - 0.3$. The extracted $T_{C}$ values are in good agreement with prior work \cite{Goral1982,Okimoto1995}. We also determine the effective $T_{C}$ inhomogeneity as the ratio of the full-width at half-maximum ($\Delta T_C$) of the peaks in Fig.~\ref{fig:Magn} (b) with respect to $T_C$ and plot this in the inset to Fig.~\ref{fig:Magn} (b). A clear increase in $\Delta T_C$/$T_C$ is observed with substitution, which could include chemical as well as true magnetic contributions to the inhomogeneity. We discuss this further in Section~\ref{section:MuSR} where we explicitly determine the magnetic inhomogeneity using $\mu$SR. The $T_{C}$ and $\Delta T_C$ values determined here are also displayed in Table~\ref{tab:table2}.

Figure~\ref{fig:Magn}(c) shows the magnetic field dependence of the magnetic moment of Y$_{1-x}$La$_{x}$TiO$_{3}$ for a range of La concentrations at 6 K. The samples measured here are the same as those in Figs.~\ref{fig:Magn}(a,b). The magnetic field was applied along the crystallographic $c$-axis. 
In the substitution range studied here, Y$_{1-x}$La$_{x}$TiO$_{3}$ is known to have zero remanence \cite{Zhou2005} (see inset to Fig.~\ref{fig:Magn}(d) for YTiO$_3$ data). We thus use an exponential extrapolation, which is known to be applicable to soft ferromagnetic materials \cite{Umenei2011}, to extract the saturation magnetic moment, which is then compared to the 7 T values in Fig.~\ref{fig:Magn}(d). 
YTiO$_{3}$ exhibits a saturation magnetization of 0.97(2) $\mu$$_{\text{B}}$/Ti, in excellent agreement with the expected spin-1/2 moment of 1 $\mu$$_{\text{B}}$ \cite{Mochizuki2004}. This is also consistent with the conclusion that YTiO$_{3}$ exhibits a completely quenched orbital moment \cite{Ulrich2002}. Note that previous works on single crystals reported substantially lower saturation moments of $\leq$ 0.84 $\mu$$_{\text{B}}$, perhaps due to oxygen off-stoichiometry \cite{Garrett1981,Tsubota2000,Ulrich2002,Kovaleva2007}. Upon La substitution, the saturation magnetization is seen to decrease rapidly: from 0.97(2) $\mu$$_{\text{B}}$/Ti at $x = 0$, to 0.63(1) $\mu$$_{\text{B}}$/Ti at $x = 0.1$, to 0.55(1) $\mu$$_{\text{B}}$/Ti at $x = 0.15$, and to 0.40(1) $\mu$$_{\text{B}}$/Ti at $x = 0.2$. With further La substitution, the decrease in saturation magnetization is weaker, reaching 0.34(1) $\mu$$_{\text{B}}$/Ti at $x = 0.25$ and 0.29(1) $\mu$$_{\text{B}}$/Ti at $x = 0.3$. These results are summarized in Table~\ref{tab:table1}.
Whereas YTiO$_{3}$ approaches almost complete saturation at 7 T, the actual saturation field steadily increases up to $x = 0.3$. Note that we have not considered here a possible paramagnetic contribution to the field-dependence of the magnetic moments observed in Fig.~\ref{fig:Magn}(c), which we discuss in Section~\ref{section:MuSR}. Such a paramagnetic component would also contribute to the exponential-form of the high-field magnetization, based on the Langevin theory of paramagnetism.

\subsection{X-Ray Magnetic Circular Dichroism}
\label{section:xmcd}
\begin{figure}
\includegraphics[width=0.45\textwidth]{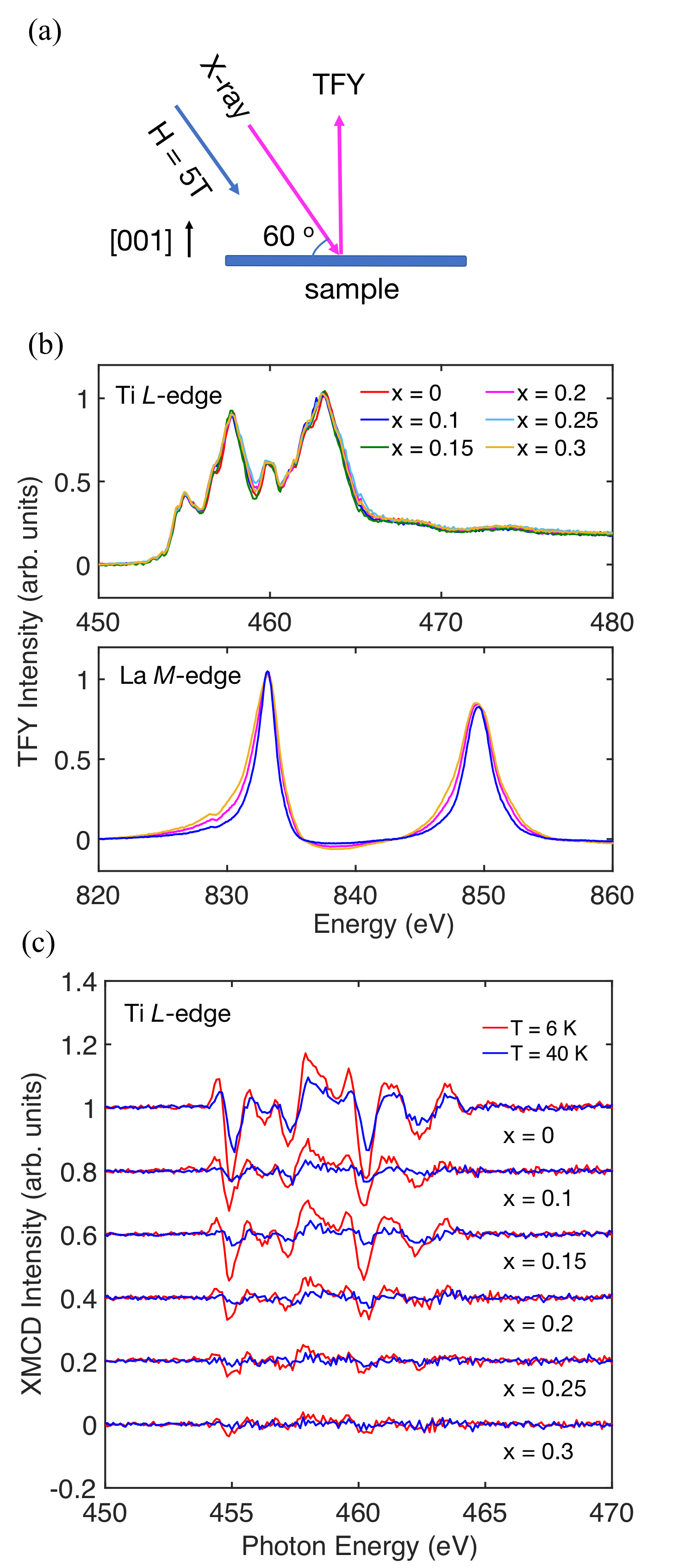}
\caption{(a) Scattering geometry for the XAS/XMCD measurements. Crystals were polished on the (001) plane, and x-rays were incident at an angle of $60^{o}$ to the sample surface. A magnetic field of 5 T was applied along the incident beam. (b) Ti $L_{2,3}$-edge XAS (top) and La $M$-edge XAS (bottom) for all the La-substitution levels at 6 K. The data measured at the lowest energy were subtracted, and the net intensity was subsequently normalized at the highest peak. (c) XMCD intensities at Ti $L_{2,3}$-edge at 6 K (FM phase) and 40 K (paramagnetic phase). The data were normalized by the intensity of the largest Ti $L_{2,3}$-edge XAS peak. The results at $x = 0, 0.1, 0.15, 0.2$ and 0.25 are shifted by 1, 0.8, 0.6, 0.4 and 0.2 arb. units, respectively, for better visibility.}
\label{fig:XMCDTdep}
\end{figure}

In order to confirm that the suppression of ordered magnetic moment with substitution arises from Ti$^{3+}$ spins, we performed XAS/XMCD measurements. 
The samples measured here are the same as those used for the magnetometry measurements. 
As shown in Fig.~\ref{fig:XMCDTdep} (a), the x-rays were incident at an angle of $60^{o}$ relative to the sample surface. The crystals were polished on the (001) plane, and a 5 T field was applied along the incident beam. Left- and right-circularly-polarized soft x-rays were tuned to the Ti $L_{2,3}$- and La $M$-edges and recorded in TFY detection mode. Figure~\ref{fig:XMCDTdep} (b) shows the XAS spectra obtained at the Ti $L_{2,3}$-edge (top) and La $M$-edge (bottom). 
The lack of an observable change in Ti XAS indicates that the oxygen stoichiometry does not change in the studied substitution range, as this would be reflected as shifts in the Ti $L$-edge peaks due to conversion from Ti$^{3+}$. 
The systematic change in the La $M$-edge response indicates the expected systematic increase with La substitution. 
It is known that high-energy x-rays tend to oxidize the surface of RTiO$_{3}$ samples, converting Ti$^{3+}$ to Ti$^{4+}$ \cite{Cao2015}. However, since TFY is a bulk-sensitive detection mode, this oxidation does not affect the measurements. The field-cooled and zero-field cooled measurements did not show any difference (data not shown here), as all the XMCD measurements were performed in a 5-T field, which is high enough to nearly saturate the magnetic moments, as seen from Fig.~\ref{fig:Magn}(c). 
Hence, the samples were always measured after zero-field cooling.

\begin{figure}
\includegraphics[width=0.4\textwidth]{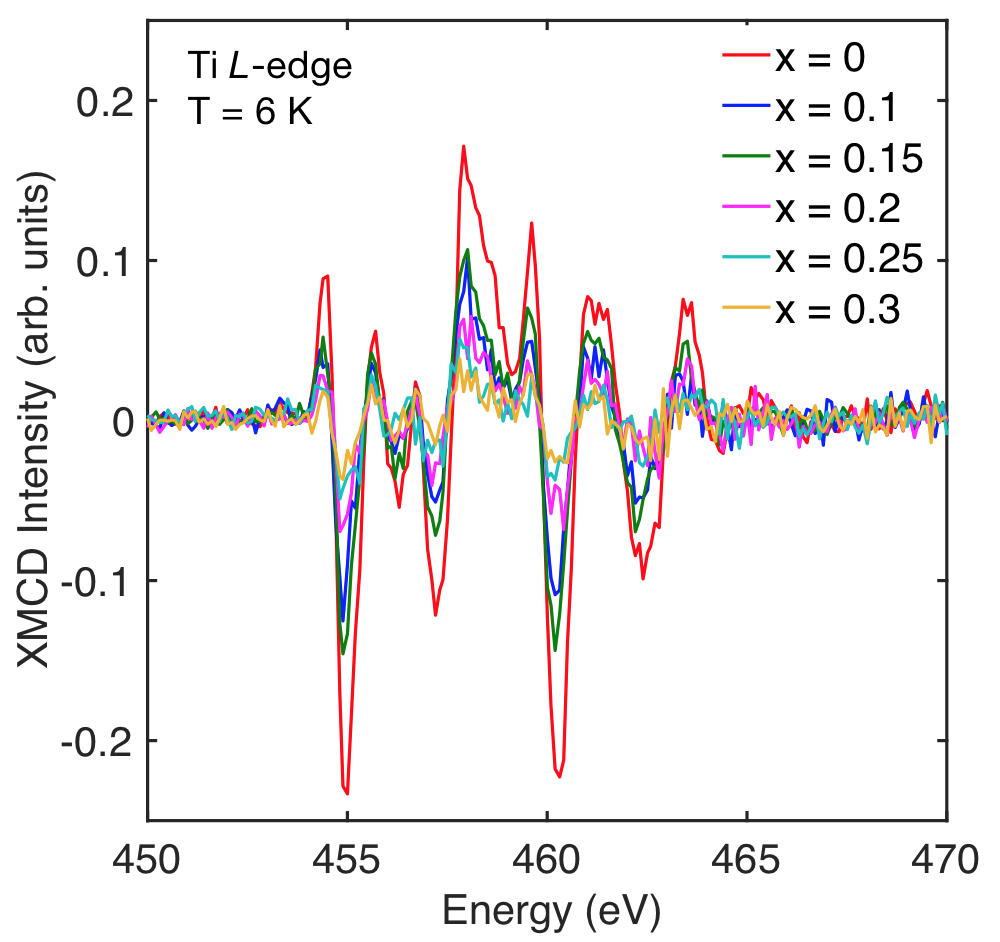}
\caption{La-substitution dependence of Ti $L_{2,3}$-edge XMCD in the FM phase at $T = 6$ K. The data are normalized by the intensity of the largest Ti $L_{2,3}$-edge XAS peak.}
\label{fig:XMCDOrder}
\end{figure}

Figure \ref{fig:XMCDTdep} (c) shows the XMCD intensities obtained at the Ti $L_{2,3}$-edge at $T = 6$ K and 40 K. As expected, the dichroism signal is stronger at 6 K, in the FM state, than at 40 K, in the paramagnetic state. As highlighted in Fig.~\ref{fig:XMCDOrder}, the XMCD signal at 6 K is strongly suppressed with substitution, indicative of a substantial weakening of the FM order. The absence of significant changes in the Ti XAS (Fig.~\ref{fig:XMCDTdep} (b, top)) indicates that the XMCD intensity suppression is intrinsic to Ti$^{3+}$ moments, and not the result of oxidation of Ti$^{3+}$ to Ti$^{4+}$; the latter is not expected to occur in an isovalently-substituted system. 
It is known that the intensity in XMCD arises from both spin and orbital moments \cite{Thole1992,Carra1993}. In principle, it is possible to extract the absolute spin and orbital moments from XMCD data through the application of sum rules \cite{Maps1995}, yet the RTiO$_3$ $L_{2,3}$-edge peaks are insufficiently separated (see Fig.~\ref{fig:XMCDTdep} (b, top)), which makes it difficult to reliably perform this estimate. However, considering that the XMCD intensities show a strong decrease above $T_C$, it is likely that the predominant contribution arises from the spin magnetic moment. This is indeed the case for YTiO$_{3}$, as we observe the full theoretically expected spin-1/2 moment of $\sim$ 1 $\mu$$_{\text{B}}$/Ti with magnetometry (Section~\ref{section:magnetometry}). Prior neutron scattering studies of YTiO$_{3}$ and LaTiO$_{3}$ have shown that the orbital moment is completely quenched in both materials, and therefore it is likely that this is also the case for the solid solution Y$_{1-x}$La$_{x}$TiO$_3$ \cite{Lynn2002,Ulrich2002}. 

\begin{figure}
\includegraphics[width=0.4\textwidth]{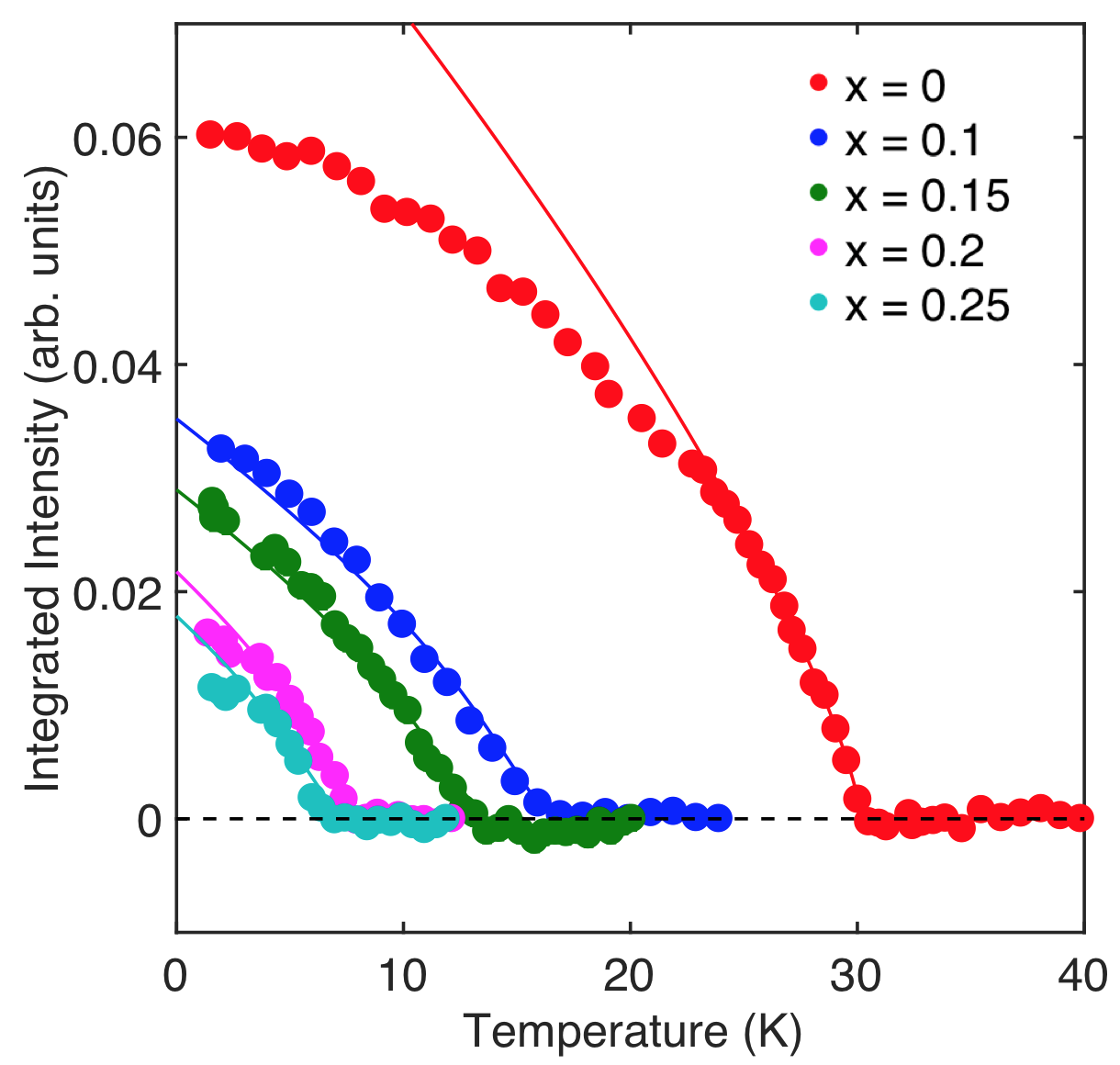}
\caption{Temperature dependences of the G-type AFM $(011)_{o}$ Bragg intensity. For $x = 0$, the line indicates a power-law fit to the order-parameter form $I \propto (1-T/T_{C})^{2\beta}$; for $x > 0$, the lines are the results of fits to a convolution of this power-law with a Gaussian distribution of $T_{C}$, as described in the text.}
\label{fig:NeutronAFM}
\end{figure}

\begin{figure}
\includegraphics[width=0.4\textwidth]{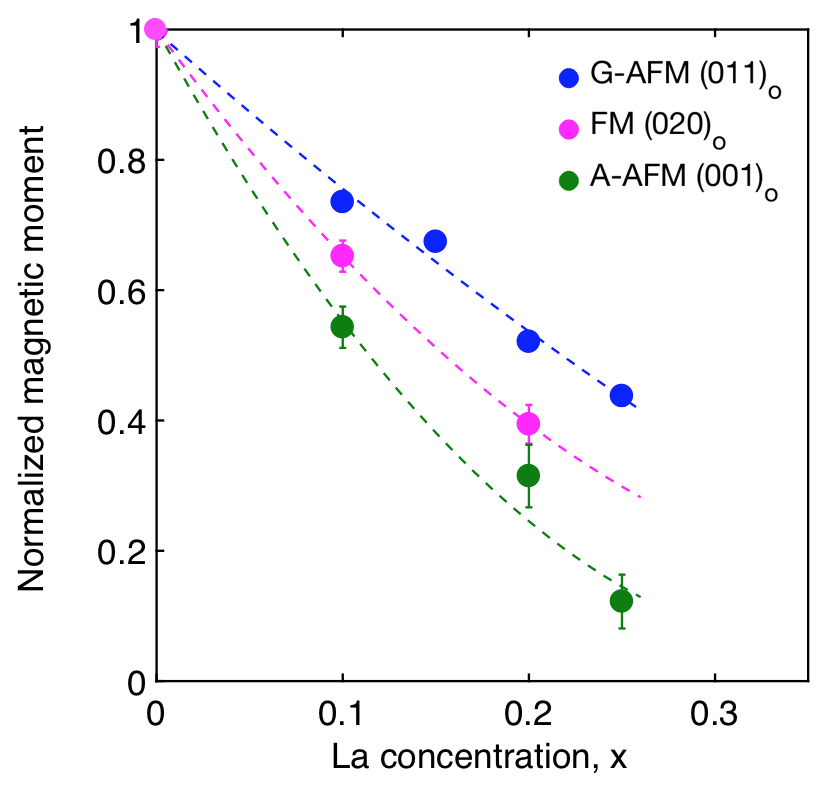}
\caption{La-concentration dependences of the FM, G-type AFM, and A-type AFM moments at $T = 2$ K obtained from the intensities of the respective magnetic Bragg reflections $(020)_{\text{o}}$, $(011)_{\text{o}}$, and $(001)_{\text{o}}$. The data are normalized to the $x = 0$ values and hence overlap at $x = 0$. Dashed lines are guides to the eye obtained from second-order polynomial fits.}
\label{fig:neutronmoments}
\end{figure}


\subsection{Neutron Diffraction}
\label{section:neutron}

Apart from the predominant FM ordered moment along the $c$-axis, weak G-type and A-type AFM moments are found in the ordered state of YTiO$_3$ along the $a$-axis and $b$-axis, respectively, due to a canting of spins \cite{Ulrich2002}. The FM Bragg reflections coincide with strong nuclear Bragg peaks, which would require long counting times to accurately extract the magnetic component. Therefore, we use instead the $(011)_{\text{o}}$ G-type AFM Bragg reflection. Here, the subscript ``o'' refers to the orthorhombic notation.
Figure~\ref{fig:NeutronAFM} shows the temperature and La-substitution dependence of the $(011)_{\text{o}}$ G-type AFM Bragg reflection. 
A (relatively weak) temperature-independent nuclear contribution was subtracted and the net intensity was normalized by the $(020)_{\text{o}}$ nuclear Bragg reflection intensity. This normalization was found to be consistent with normalization by sample mass, and is hence unaffected by extinction. 
For $x=0$, a power-law fit of the form $I \propto (1-T/T_{C})^{2\beta}$ in the vicinity of $T_{C}$ gives a critical exponent estimate $\beta = 0.36(4)$, consistent with the 3D Ising value of 0.3265(7) \cite{Odor2004}, which is expected due to the system's 3D Ising anisotropy \cite{Tsubota2000,Kovaleva2007} (see also Section~\ref{section:easyaxis}).
A similar determination of $\beta$ for the La-substituted crystals was not possible due to the chemical/magnetic inhomogeneity-induced smearing of the transition. Instead, we fixed $\beta$ at 0.35 (average of all 3D Ising, XY and Heisenberg exponents) and fit the data by convolving the power-law form with a Gaussian distribution of $T_{C}$. The latter yielded an estimate of the effective $T_{C}$ inhomogeneity. The solid lines in Fig.~\ref{fig:NeutronAFM} correspond to these fits. Table~\ref{tab:table2} displays the extracted $T_{C}$ as well as the full-width at half-maximum ($\Delta T_{C}$) of the Gaussian distribution of $T_{C}$. If we quantify the $T_{C}$ inhomogeneity as $\Delta T_{C}/T_{C}$, we can clearly see from Table~\ref{tab:table2} that the effective $T_{C}$ inhomogeneity increases from $\sim33\%$ at $x = 0.1$ to $\sim59\%$ at $x = 0.25$. This could include both chemical as well as true magnetic contributions to the inhomogeneity, as discussed in Section~\ref{section:MuSR}. Note that the thermal phase transitions are likely first-order, as discussed in Sections~\ref{section:MuSR} and ~\ref{section:sans}. Therefore, the analysis carried out here must be taken as merely a description of the data.
A comparison of the extracted $T_{C}$ with those obtained from other probes will be made in Section~\ref{section:MuSR}.
The intensity of the $(011)_{\text{o}}$ reflection, being proportional to the square of the staggered magnetic moment, clearly shows a suppression of the AFM ordered moment with increasing La concentration.

Figure~\ref{fig:neutronmoments} compares the La concentration dependences of the FM, G-type AFM, and A-type AFM moments obtained from the intensities of the respective Bragg reflections $(020)_{\text{o}}$, $(011)_{\text{o}}$, and $(001)_{\text{o}}$. 
A clear suppression of all three ordered magnetic components is observed. 
In Section~\ref{section:easyaxis}, we discuss the La concentration dependence of the magneto-crystalline anisotropy and its implications for the neutron diffraction results presented here.

\subsection{Comparison of Magnetic Moment Results}
\label{section:comparison}

\begin{figure}
\includegraphics[width=0.45\textwidth]{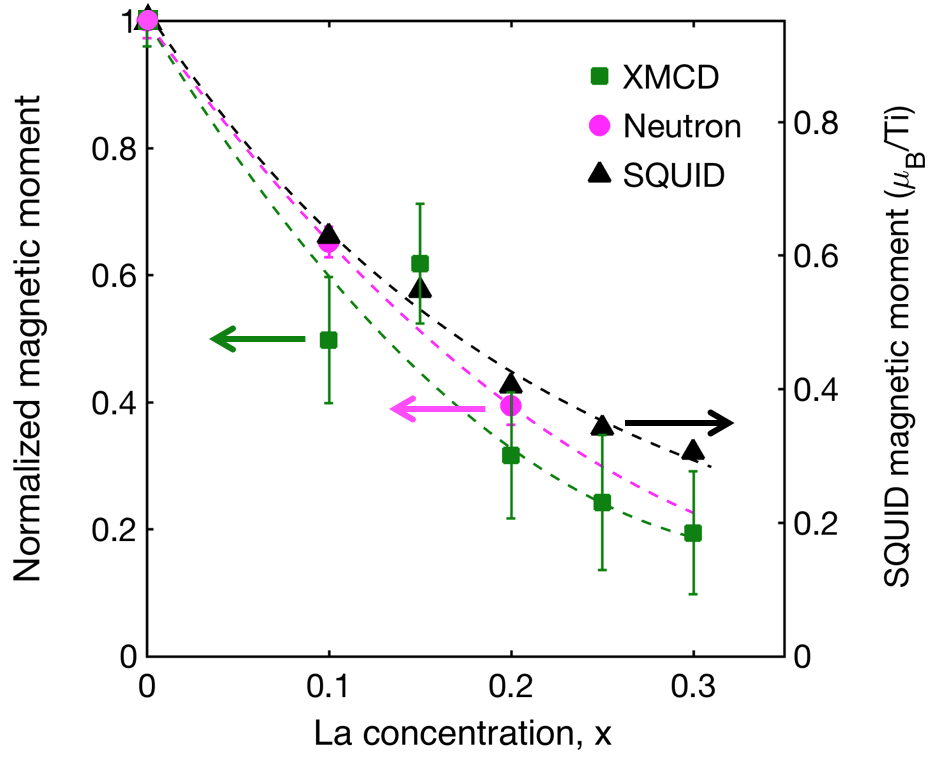}
\caption{La-concentration dependences of the FM moment obtained from energy-integrated absolute XMCD intensities at 6 K and neutron diffraction intensities of the FM Bragg reflection $(020)_{\text{o}}$ at 2 K (left vertical axis). The neutron and x-ray data are normalized to the respective $x = 0$ values. The absolute FM moment obtained from the extrapolated SQUID magnetometry data at 6 K is added for comparison (right vertical axis). Dashed lines are guides to the eye obtained from second-order polynomial fits.}
\label{fig:FMmoments}
\end{figure}
 
Figure~\ref{fig:FMmoments} compares the La concentration dependences of the FM moment obtained from the extrapolated SQUID magnetometry data and XMCD (both at 6 K) with the neutron diffraction results (at 2 K). As indicated in Section~\ref{section:xmcd}, it is rather difficult to estimate the spin magnetic moment from the XMCD data. Nevertheless, we use the La-concentration dependence of the energy-integrated absolute XMCD intensities in Fig~\ref{fig:FMmoments}. The FM moments obtained from neutron diffraction and SQUID magnetometry show excellent agreement. The ordered moments obtained from XMCD have larger uncertainty, yet also show good agreement, which indicates that the suppression of the spin magnetic moment is likely simply reflected as a scale factor in the XMCD intensities. Note that this assumes a negligible contribution from the orbital moment, as explained in Section~\ref{section:xmcd}.
 
The results discussed so far clearly indicate a significant decrease of the ordered magnetic moment in Y$_{1-x}$La$_{x}$TiO$_{3}$ with increasing La concentration. However, the data from these bulk probes do not take into account a possible reduction of the magnetic volume fraction below 100\%. In the next Section, we describe detailed complementary $\mu$SR measurements aimed to distinguish between a decrease in uniform local moment and magnetic volume fraction.


\subsection{Muon Spin Rotation}
\label{section:MuSR}

\begin{figure}
\includegraphics[width=0.45\textwidth]{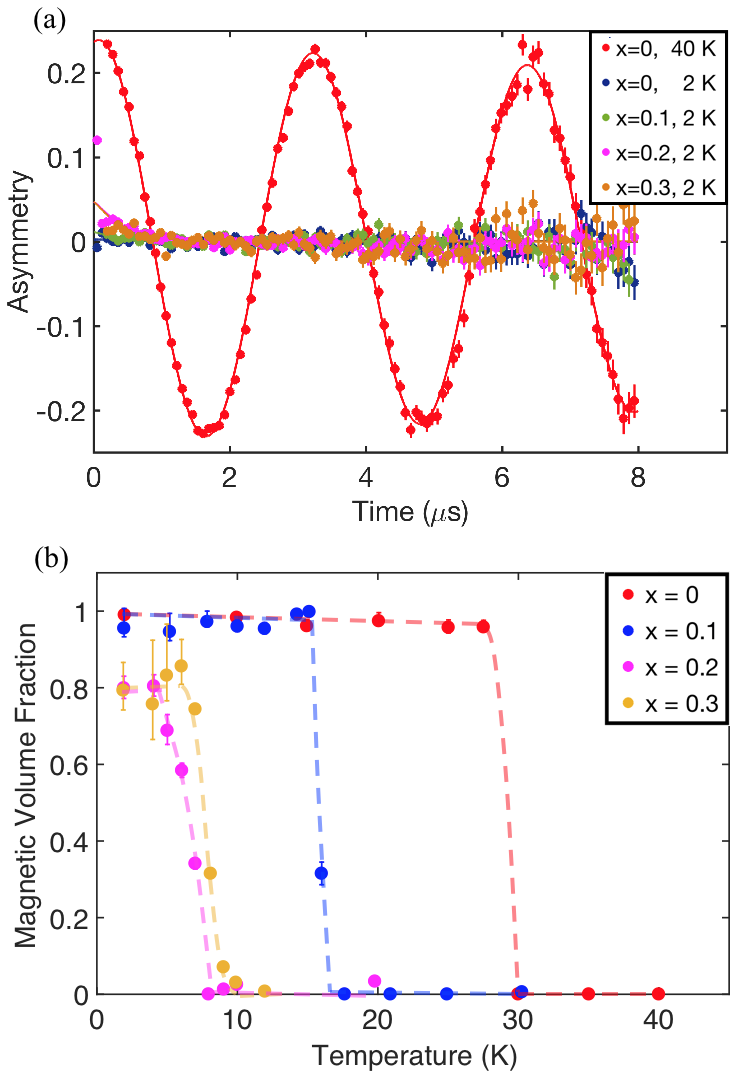}
\caption{(a) wTF $\mu$SR spectra for $x$ = 0, 0.1, 0.2 and 0.3 in the ordered phase at $T$ = 2 K and for $x$ = 0 in the paramagnetic phase at $T$ = 40 K for comparison. Solid lines are fits to the data at each temperature, as described in the text. (b) Temperature and La-concentration dependence of the magnetic volume fraction in Y$_{1-x}$La$_{x}$TiO$_{3}$. Dashed lines are guides to the eye.}
\label{fig:wTFPlots}
\end{figure}

\begin{figure*}
\includegraphics[width=0.95\textwidth]{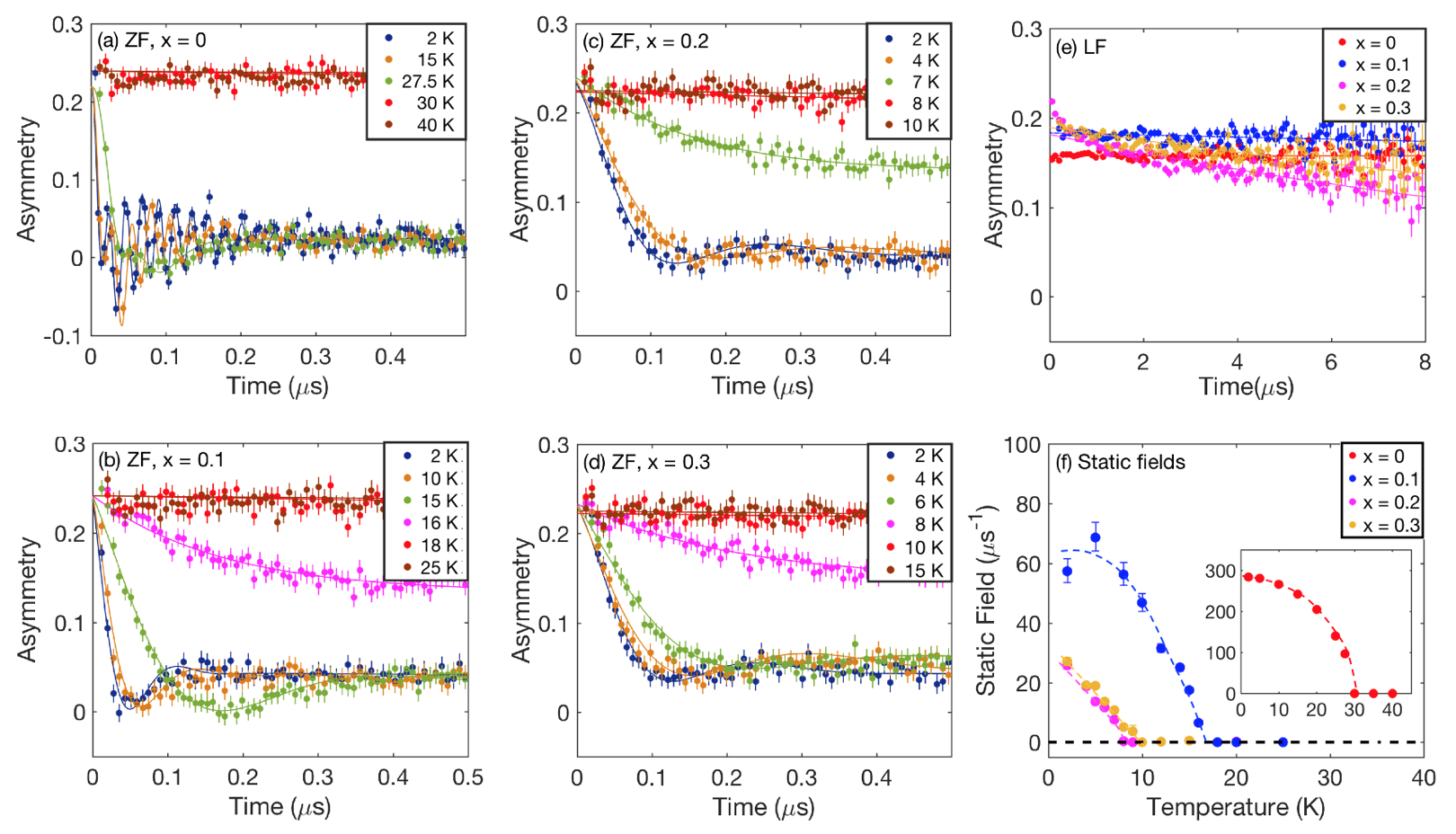}
\caption{(a-d) ZF $\mu$SR spectra at different temperatures for (a) $x$ = 0, (b) $x$ = 0.1, (c) $x$ = 0.2 and (d) $x$ = 0.3. Solid lines are fits to the data at each temperature, as described in the text. (e) LF $\mu$SR spectra at 2 K. The fields applied were 2 kG for $x$ = 0, 1.5 kG for $x$ = 0.1 and 1 kG for $x$ = 0.2 and 0.3. (f) Static local fields as a function of temperature and La concentration obtained as described in the text. For clarity, the $x$ = 0 data are shown separately as an inset. The dashed lines are guides to the eye.}
\label{fig:ZFPlots}
\end{figure*}

$\mu$SR is a highly sensitive probe of magnetism that can independently measure the (local) magnetic order parameter and the magnetically-ordered volume fraction. In $\mu$SR, positively-charged spin-polarized muons are implanted in the sample and generally stop at an interstitial position within the crystal, without significant loss of polarization. The subsequent polarization of the muon is determined by its local magnetic environment, which is measured via the asymmetric decay of the muon. $\mu$SR experiments have previously been carried out on the end members YTiO$_{3}$ and LaTiO$_{3}$ and revealed the magnetically-ordered ground state in these systems \cite{Baker2008}. Here, we report $\mu$SR in zero applied magnetic field (ZF), weak transverse field (wTF), and longitudinal field (LF), at temperatures that span the FM-paramagnetic phase transition in the same samples as those used for the neutron diffraction measurements. 
The spectra are analyzed in the time domain using least-squares minimization routines in the MusrFit software package \cite{Suter2012}. 

wTF $\mu$SR measurements give direct information about the magnetic volume fraction. A spectrum with no oscillations indicates a completely ordered sample, whereas a spectrum with undamped oscillations indicates a fully disordered sample. Figure~\ref{fig:wTFPlots}(a) shows the wTF spectra for four substitution levels spanning the range $0 \leq x \leq 0.3$. The spectra are fit to the function

\begin{equation}
A(t) = a_{\text{p}}{e}^{-\lambda t}\cos{(\omega t + \phi),}
\label{eq:1}
\end{equation}
where $a_{p}$ is the amplitude of the oscillating component, which is directly proportional to the paramagnetic volume fraction, $\lambda$ is the damping due to paramagnetic spin fluctuations, $\omega$ is the Larmor precession frequency, which is set by the strength of the transverse magnetic field, and $\phi$ is the phase offset. The zero for $A(t)$ was allowed to vary in order to account for the asymmetry baseline shift that is known to occur in magnetically-ordered samples. As expected, the 40 K spectra for YTiO$_{3}$ exhibit full-amplitude oscillations indicative of a paramagnetic phase.  The result for the magnetic volume fraction, $F = 1 - a_{p}(T)/a_{p}(T>T_{C})$, is shown in Fig.~\ref{fig:wTFPlots}(b). For $x$ = 0 and $x$ = 0.1, we observe completely ordered magnetic volumes at low temperature, whereas the $x$ = 0.2 and 0.3 data reveal phase separation into paramagnetic and magnetically-ordered regions, with respective low-temperature volume fractions of  $\sim20$\% and $\sim80$\% for both La concentrations.

In order to determine the temperature dependence of the local ordered moments, ZF-$\mu$SR spectra were acquired. Figure~\ref{fig:ZFPlots}(a)-(d) shows the ZF-$\mu$SR time series spectra measured for $x$ = 0, 0.1, 0.2 and 0.3. Clear oscillations indicative of magnetic order are visible for each sample below the respective $T_C$. Above the transition temperature, the spectra exhibit exponential relaxation characteristic of the paramagnetic disordered phase. In the ordered phase, the ZF spectra are well described by the function

\begin{equation}
\begin{split}
A(t) = F\bigg(\sum^{n}_{i=1}s_{i}r_{i} \mathrm{e}^{-\lambda_{i}^{\text{T}}t^2}\cos{(2\pi \nu_{i}t + \phi_{i})}
\\+~s_{i}(1-r_{i}) \mathrm{e}^{-\lambda^{\text{L}}t} \bigg)+~(1-F)~a_{\text{p}} \mathrm{e}^{-\lambda_{\text{p}}t}.
\label{eq:2}
\end{split}
\end{equation}
Here, $F$ is the ordered magnetic fraction characterized by two components: an oscillating transverse component with a precession frequency $\nu_{i}$, phase offset $\phi_{i}$ and damping rate $\lambda_{i}^{\text{T}}$ due to a distribution of the local fields; a slowly relaxing longitudinal component due to a parallel orientation of the local field and muon spin polarization. In polycrystalline samples, the local fields are randomly oriented, and the longitudinal part gives rise to the usual \textquotedblleft1/3\textquotedblright component with $1 - r_{i}=1/3$. In single crystals, however, $r_{i}$ depends on the local field direction and varies between 0 and 1 as the orientation of the field with respect to the muon polarization direction changes from parallel to perpendicular. An additional non-oscillating component arises from the paramagnetic volume fraction $(1-F)$, with the corresponding relaxation rate $\lambda_{\text{p}}$. For the fits to the ZF data, $F$ was fixed based on the wTF results from Fig.~\ref{fig:wTFPlots}. The spectra for $x$ = 0 were best modeled with two muon stoppage sites ($n$ = 2), similar to Ref. \cite{Baker2008}. The site ratio $s_{1}/(s_{1} + s_{2})$ was fixed at 0.314 based on the best fit obtained at 2 K. The ratio of the precession frequencies was found to be $\sim$ 0.2 for the two muon stoppage sites. The data for the La-substituted samples were well-fit with a single muon stoppage site ($n$ = 1), likely because the site corresponding to the lower precession frequency has too small a frequency to be resolved in these samples. Further, for the La-substituted samples, the Gaussian functions in the first term on the right hand side of Eq. \eqref{eq:2} had to be replaced by exponentials in order to obtain a good fit, indicating a change in the local field distribution likely due to disorder from La-substitution. Note that, at all substitution levels, the spectra above $T_C$ are free of oscillations, as expected, and well fit by a single exponentially-relaxing component, which is typical for paramagnets. The application of a small longitudinal field between 1 kG and 2 kG substantially decouples the ZF relaxation at 2 K, as shown in Fig.~\ref{fig:ZFPlots}(e), which confirms that the ZF relaxation originates from static magnetic order. The magnitude of the static internal field (which is proportional to the local ordered moment) can be calculated as $B_{\text{int}}$ = $\sqrt{(2\pi\nu)^2 + (\lambda^{T})^2}$, where $\nu$ is the maximum precession frequency and $\lambda^{T}$ is the corresponding relaxation rate. This includes both homogeneous and inhomogeneous contributions that arise, respectively, from the well-defined average local field and the damping due to the distribution of local fields around the average value. Figure~\ref{fig:ZFPlots}(f) shows the static internal fields thus obtained. A clear suppression of the local ordered moment can be seen. We therefore arrive at the conclusion that the strong suppression of the bulk ordered moment occurs due to a \textit{combination} of reduced magnetic volume fraction and a suppressed local ordered moment. This is in contrast to systems such as Fe$_{1.03}$Te, for example, in which a pressure-induced AFM-FM transition occurs through a weakening of local AFM moment while maintaining a 100\% AFM volume fraction up to the vicinity of the phase boundary \cite{Bendele2013}.

\begin{figure}
\includegraphics[width=0.45\textwidth]{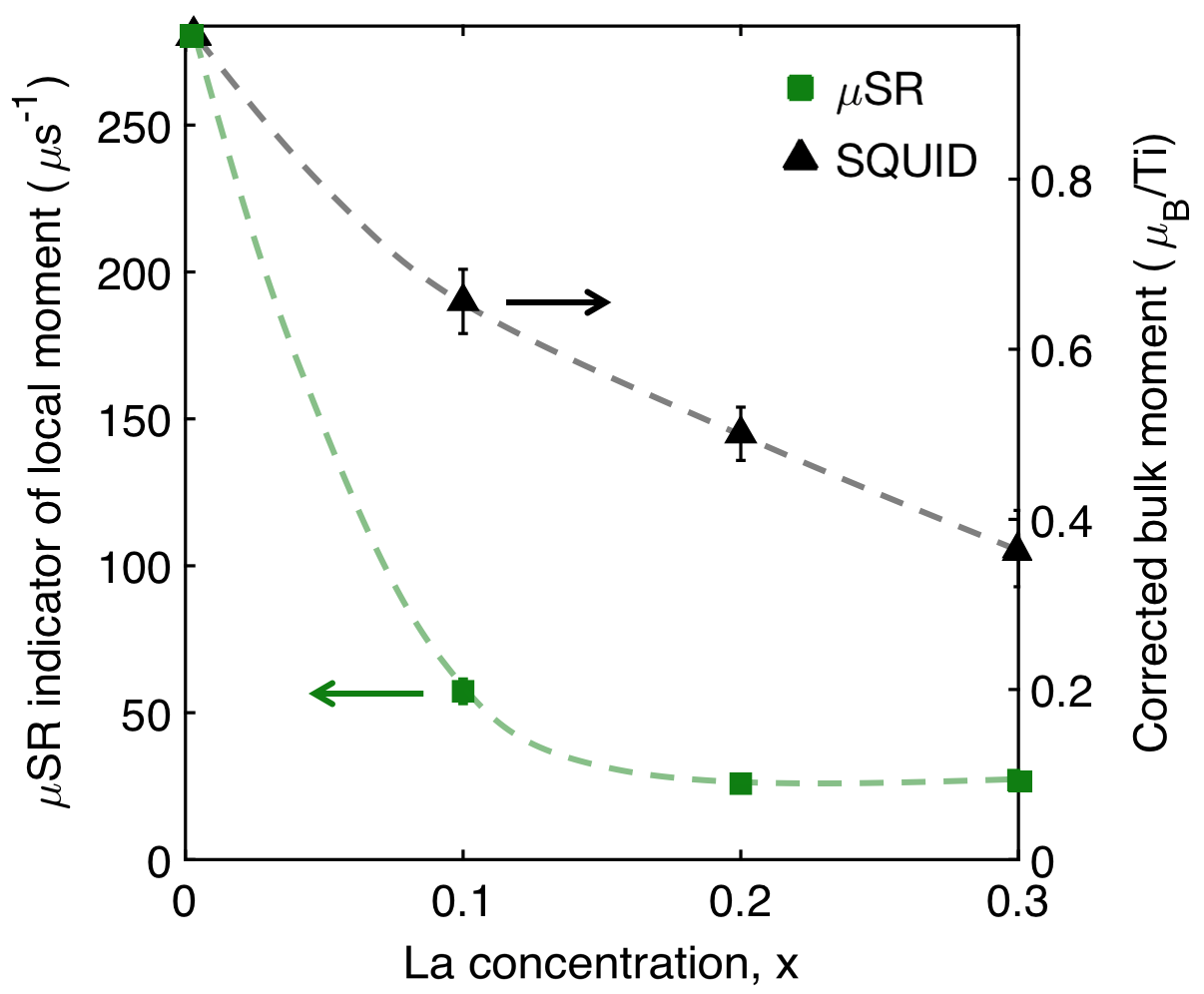}
\caption{La-concentration dependences of the local magnetic moment obtained from $\mu$SR (left vertical axis) and from the extrapolated bulk SQUID FM moment corrected for the volume fraction (right vertical axis). The vertical axes are scaled such that the data overlap at $x = 0$, allowing for a direct comparison. Dashed lines are guides to the eye.}
\label{fig:LocalvsBulk}
\end{figure}

Although $\mu$SR cannot distinguish between FM and AFM order, the refinement of the ZF spectra indicates the presence of only one precessing component in both $x$ = 0.2 and 0.3. This is strong evidence against phase separation into FM and AFM regions, as this would lead to the presence of at least two precessing components due to two magnetically inequivalent muon stoppage sites \cite{Bendele2013}. 

We can now determine the bulk magnetic-moment corrected for the reduced volume fraction for $x = 0.2$ and 0.3. Figure~\ref{fig:LocalvsBulk} compares the extrapolated bulk SQUID FM moment, corrected for the volume fraction, with the local moment obtained from $\mu$SR. While the suppression of ordered moments is clearly observed in both, the local moment obtained from $\mu$SR decays much more strongly. Dipole-field calculations indicate that the muon stoppage sites in YTiO$_{3}$ are close to the Ti$^{3+}$ ions \cite{Baker2008}. Therefore, small changes in the canting angle (and magneto-crystalline anisotropy, as described in Sections~\ref{section:sans} and~\ref{section:easyaxis}) may significantly alter the muon local magnetic environment, rendering a direct comparison of the $\mu$SR local moments at different substitution levels difficult. 

\begin{figure}
\includegraphics[width=0.45\textwidth]{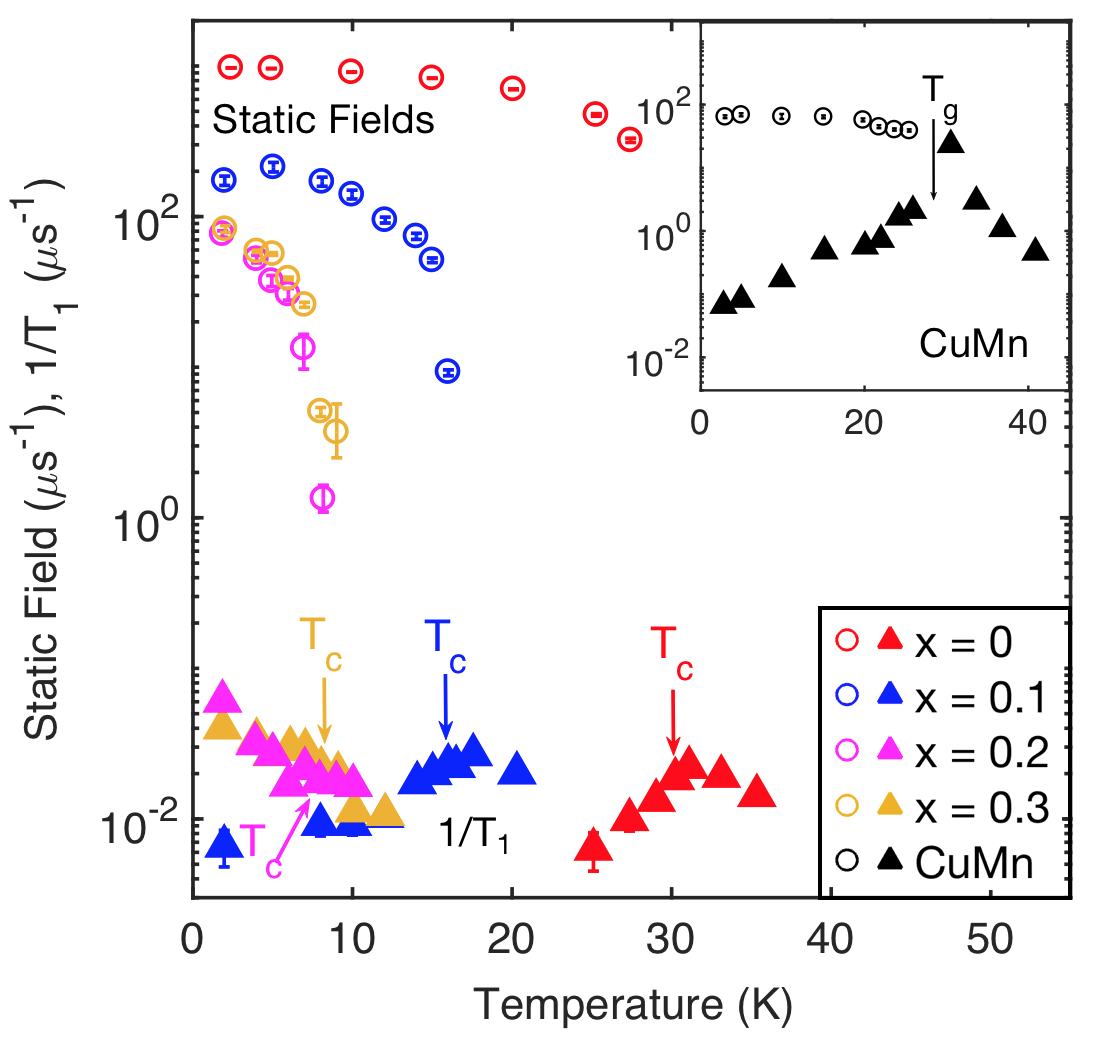}
\caption{Static internal fields $B_{\text{int}}$ (open circles) and longitudinal relaxation rates $1/T_1$ (closed triangles) in Y$_{1-x}$La$_{x}$TiO$_{3}$. The inset shows the case of the spin-glass system CuMn (taken from Ref.\cite{Uemura1985}).}
\label{fig:LFrate}
\end{figure}

\begin{figure}
\includegraphics[width=0.4\textwidth]{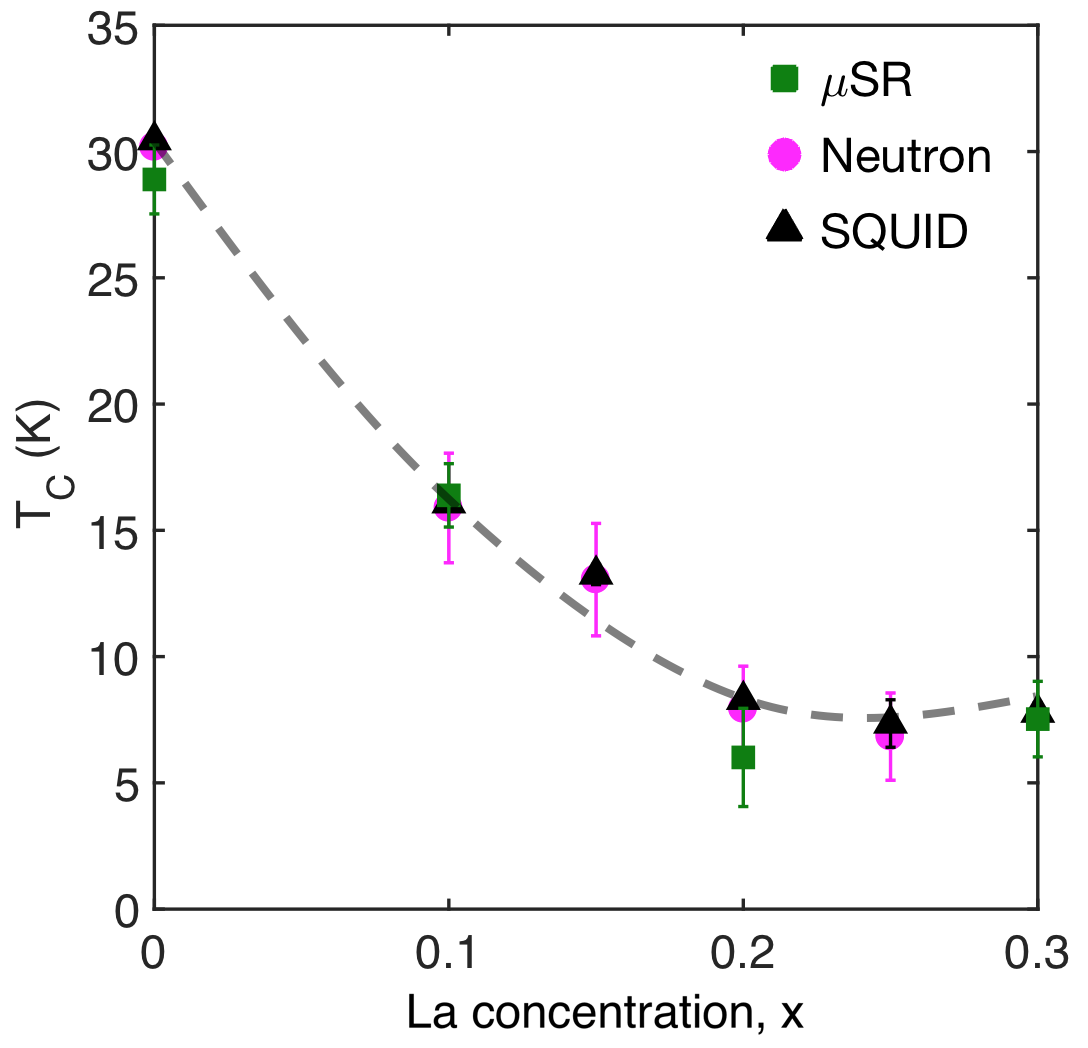}
\caption{Comparison of the magnetic transition temperatures at different substitution levels, obtained with different probes. Dashed line is a guide to the eye.}
\label{fig:TcComp}
\end{figure}

The first- versus second-order nature of a magnetic phase transition can be discerned from the dynamic critical behavior of a system. To this end, we also performed measurements at temperatures spanning the FM-paramagnetic transition in a small longitudinal field (LF) just sufficient to decouple the 2 K ZF relaxation by $\sim$70\%. The muon spin is then parallel to the external field, and the internal field depolarizes the muon spin with a muon depolarization rate (decay rate) equivalent to the spin-lattice relaxation (1/T$_1$). The LF field data were therefore fit to the form

\begin{equation}
A(t) = a_{\text{L}}{e}^{-t/T_{1}}.
\label{eq:3}
\end{equation}
The applied LF fields were 2 kG for $x$ = 0, 1.5 kG for $x$ = 0.1, and 1 kG for $x$ = 0.2 and 0.3. Figure~\ref{fig:LFrate} shows the temperature dependence of the dynamic relaxation rate $1/T_{1}$ along with the static internal fields. The transition temperatures displayed are extracted from Fig.~\ref{fig:ZFPlots}(f) as the onset of a non-zero static internal field. Whereas a peak-like feature is seen for $x$ = 0 and 0.1 at a temperature slightly above $T_{C}$, no clear peak is observed for $x$ = 0.2 and 0.3. 
Moreover, the magnitude of $1/T_{1}$ is less than 1\% of the static internal field, even at $x = 0$ and 0.1. 
This is in stark contrast to the case of continuous phase transitions such as in (Mn,Fe)Si \cite{Goko2017} and dilute spin glasses such as CuMn \cite{Uemura1985} (see inset of Fig.~\ref{fig:LFrate}) in which $1/T_{1}$  exhibits a clear peak with a peak value of at least $\sim$10\% of the static internal field. 
This indicates that the thermal phase transitions at $x = 0$ and 0.1 might already be weakly first-order \cite{Barsov1983}, even though the neutron diffraction data at $x = 0$ are also consistent with Ising criticality. 
The absence of a clear signature of dynamical critical behavior at $x = 0.2$ and 0.3, along with the observation of significant phase separation into paramagnetic and magnetically-ordered regions near their respective transition temperatures indicate that the thermal phase transition for $x = 0.2$ and 0.3 is likely first-order. 
This conclusion is further supported by the absence of a sharp divergence of the specific heat at $x = 0.12, 0.17, 0.23$ and 0.30, as reported previously, near the respective transition temperatures \cite{Zhao2015}. An increasing broadening of the thermal phase transition with increasing La concentration can also be observed in our magnetometry and neutron data, in the form of an increasing $\Delta T_C / T_C$ (see Table~\ref{tab:table2} and the inset to Fig.~\ref{fig:Magn}(b)).
However, we cannot fully rule out the possibility that this is the result of chemical inhomogeneity related to La substitution. 
In order to clarify this, we compare the transition temperatures obtained from the bulk SQUID and neutron measurements with those from local $\mu$SR measurements in Fig.~\ref{fig:TcComp} and Table~\ref{tab:table2}. The transition temperatures from the SQUID data are determined as the peak in $dM/dT$ in Fig.~\ref{fig:Magn} (b). The errors are estimated as the temperature at which $dM/dT$ is 95\% of its value at $T_{C}$. In the case of the neutron measurements, the transition temperatures are determined from the power-law fits in Fig.~\ref{fig:NeutronAFM} and the errors for $x>0$  are estimated as one standard deviation of the fitted Gaussian distribution of $T_{C}$. For $\mu$SR, the transition temperatures are taken as the midpoints of the transitions observed in the magnetically-ordered volume fraction in Fig.~\ref{fig:wTFPlots}, with the errors estimated as the width of the transition.
The transition temperatures obtained from the bulk and local probes are in excellent agreement, which indicates that the local and average bulk chemical compositions do not differ significantly.
This is considerable evidence against significant inhomogeneity effects in the $\mu$SR results reported here.

\begin{table}[h!]
  \begin{center}
    \caption{Magnetic transition temperatures $T_C$ and their effective distribution $\Delta T_C$ (in Kelvin) obtained from different probes.}
    \label{tab:table2}
    \vspace*{5mm}
    \begin{tabular}{l|c|c|c|c|c}
      \multirow{2}{*}{\textbf{x}} & \multicolumn{2}{|c|}{\textbf{SQUID}} & \multicolumn{2}{|c|}{\textbf{Neutron}} & \textbf{$\mu$SR}\\
      \cline{2-5}
      &  \textbf{$T_C$} &  \textbf{$\Delta T_C$} & \textbf{$T_C$} & \textbf{$\Delta T_C$}  & \textbf{$T_C$}  \\
      \hline
     0 & 30.4(0.1) & 2.3 & 30.2(0.1) & 0.2 & 28.9(1.4)\\
      0.1 & 16.1(0.3) & 5.1 & 15.9(2.2) & 5.2 & 16.4(1.3)\\
      0.15 & 13.2(0.4) & 6.6 & 13.1(2.2) & 5.2 & - \\
      0.2 & 8.3(0.3) & 5.4 & 7.9(1.7) & 4.0 & 6.0(1.9)\\
      0.25 & 7.3(0.9) & 5.7 & 6.8(1.7) & 4.0 & - \\
     0.3 & 7.8(0.4) & 4.8 & - & - & 7.5(1.5)\\
    \end{tabular}
  \end{center}
\end{table}

\subsection{Comparison between neutron diffraction and $\mu$SR}
\label{section:muSRvsNeutron}
  
The present study provides a new example of probing critical behavior of a system via both neutron scattering and $\mu$SR.  Previously, such an attempt was made for the quantum critical evolution of MnSi tuned by hydrostatic pressure. MnSi is an itinerant-electron magnet that displays helical magnetic order below $T_C \sim$ 29 K \cite{Pfleiderer1997}. Application of hydrostatic pressure turns the system into a paramagnet above a critical pressure $p_c \sim$ 15 kbar, with first-order-like response observed in the magnetic susceptibility between $p^* \sim$ 12 kbar and $p_c$ \cite{Pfleiderer1997}. For $p < p_c$, neutron scattering studies observed a satellite Bragg peak corresponding to the helical order, with a continuous-in-temperature evolution of Bragg peak intensities below $T_C$ \cite{Pfleiderer2004,Fak2005}, similar to our results for Y$_{1-x}$La$_x$TiO$_3$ in Fig.~\ref{fig:NeutronAFM}. For $p_c < p < $19 kbar, an additional diffuse quasielastic response was observed and termed ``partial order''. 

$\mu$SR studies, on the other hand, revealed that the static magnetic order in MnSi survives with a 100\% magnetic volume fraction only up to $p^*$, followed by phase separation into static-magnetically-ordered ($F$) and paramagnetic volumes ($1-F$) between $p^*$ and $p_c$ \cite{Uemura2007}.  No dynamic critical behavior was observed for $p^* < p < p_c$, with the muon spin relaxation rate $1/T_1$ below the detection limit for pressures above $p_c$.  In the case of MnSi, these observations clearly revealed that the continuous change of the Bragg-peak intensity in neutron scattering does not imply a second-order phase transition. The reason for this is that the magnetic Bragg peak intensity scales as $S^2F$, where $S$ is the size of the local ordered moment.  In general, neutron measurements cannot distinguish a continuous evolution of $S^2$ with a 100\% magnetic volume fraction from a discontinuous change of $S^2$ at $T_C$ with a continuous change of $F$ below $T_C$. In the present study, the $\mu$SR results in Fig.~\ref{fig:wTFPlots}(b) clearly show a gradual evolution of $F$ below $T_C$ for $x = 0.2$ and $0.3$.  Thus, the neutron data in Fig.~\ref{fig:NeutronAFM} for these samples should not be regarded as indicative of a second-order thermal phase transition.

For MnSi, the apparent difference between $\mu$SR (no observable $1/T_1$) and the quasi-elastic neutron signal indicates that the spin correlations above $p_c$ are dynamic, within the energy resolution of the neutron studies in ref.\cite{Pfleiderer2004}, but too fast for $\mu$SR to detect. In the two neutron studies on MnSi \cite{Pfleiderer2004,Fak2005}, the satellite Bragg peak for helical order was observed only below $p_c$ while diffusive signal in the quasi-elastic study was observed well above $p_c$ \cite{Pfleiderer2004}, which indicates a change of spatial correlations at $p_c$. These results demonstrate that, for complex systems, the comparison of neutron scattering and $\mu$SR results may need to go beyond a simple consideration of static volume fractions and ordered moment sizes, and require careful attention to dynamic time scales and spatial correlation patterns. 

The present case for YTiO$_3$ requires particularly careful attention. The $\mu$SR results for $x = 0$ in Fig.~\ref{fig:wTFPlots}(b) indicate the absence of clear phase separation below $T_C$.  The neutron diffraction results in Fig.~\ref{fig:NeutronAFM} show a continuous temperature dependence with a reasonable power-law behavior at $T_C$. Prior specific-heat measurements \cite{Zhao2015} revealed a sharp peak of the magnetic contribution below $T_C$.  These are behaviors consistent with a second-order thermal transition, although they are not inconsistent with a weakly first-order thermal transition.  On the other hand, no dynamic critical behavior in $1/T_1$ at $T_C$ is observed by $\mu$SR.  At this point, it is not possible to unambiguously conclude whether unsubstituted YTiO$_3$ exhibits a second- or first-order thermal phase transition. Further studies of dynamic and spatial critical behaviors by inelastic neutron scattering would be needed to clarify this. Nevertheless, the first-order nature of the thermal transition becomes increasingly clear with La-substitution, as inferred from the magnetic-volume fraction data for $x = 0.2$ and 0.3 in Fig.~\ref{fig:wTFPlots}(b). Finally, we note that a related situation is encountered in the pyrite-structure metallic ferromagnet CoS$_2$, which is thought to exhibit a weakly first-order FM to paramagnet transition at around 120 K \cite{Wang2004}. Modest substitutions with several elements shifts this to a clear first-order transition, however, a prime example being CoS$_{2-x}$Se$_x$, where the transition has been studied by similar neutron methods to those applied here (i.e., diffraction, SANS, etc.) \cite{Sato2003}.

\subsection{Small-Angle Neutron Scattering}
\label{section:sans}

\begin{figure}
\includegraphics[width=0.5\textwidth]{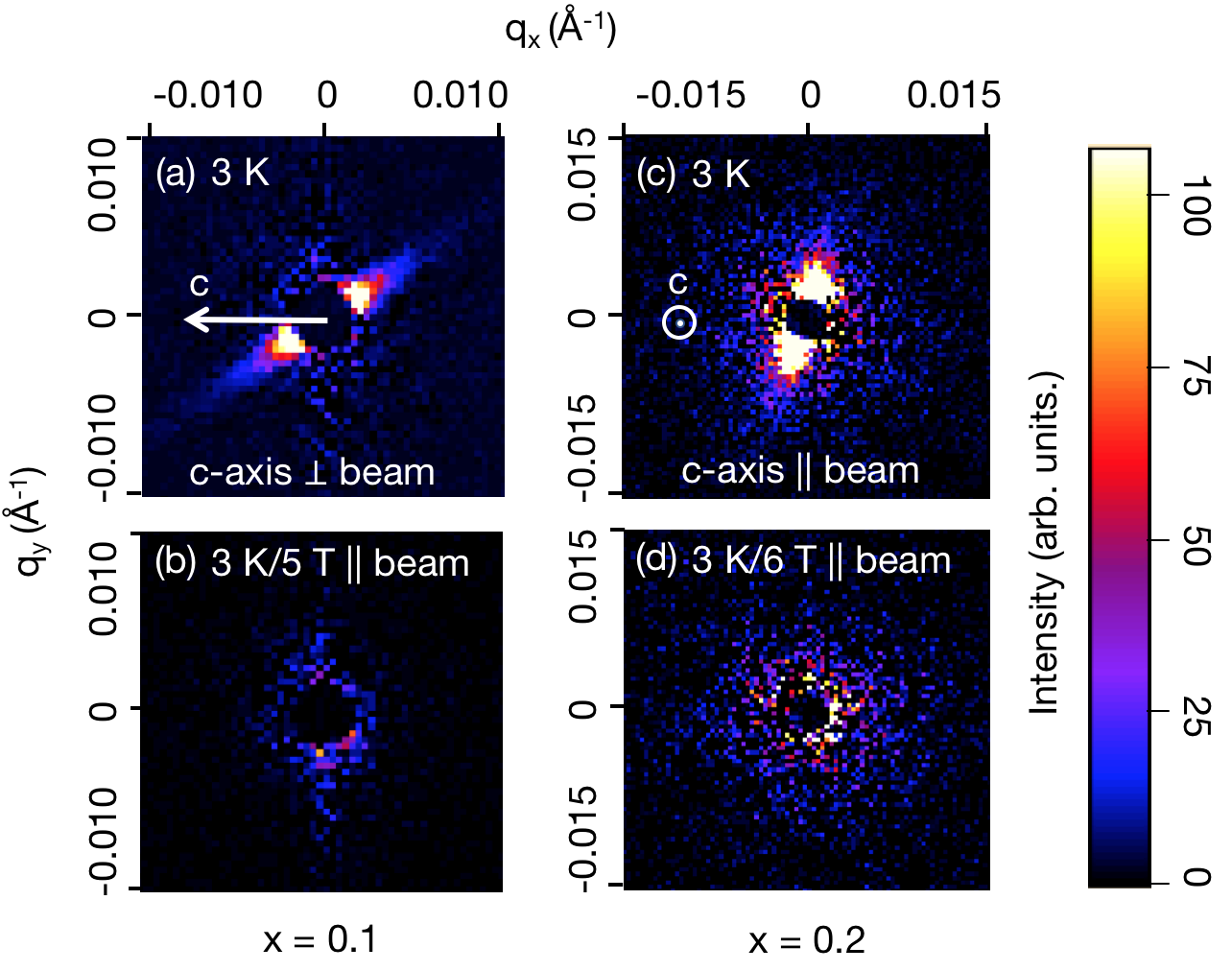}
\caption{(a,b) $q_x$-$q_y$ maps of the magnetic SANS in $x = 0.1$ at 3 K/0 T and 3 K/5 T, respectively. The magnetic field in (b) is set perpendicular to the $c$-axis of the crystal. (c,d) $q_x$-$q_y$ maps of the magnetic SANS in $x = 0.2$ at 3 K/0 T and 3 K/6 T, respectively. The magnetic field in (d) is set parallel to the $c$-axis of the crystal. The non-magnetic scattering at high temperature has been subtracted from panels (a)-(d). The black spot in the center of each image is simply due to a beam stop.}
\label{fig:SANSmaps}
\end{figure}

SANS is a premier bulk probe of magnetic ordering and inhomogeneity across nano- to mesoscopic length scales \cite{Michels2008,Sebastian2019} and is thus ideal for the study of the evolution of magnetic order with $x$ in Y$_{1-x}$La$_{x}$TiO$_3$. In particular, we use $T$-, $H$-, and $x$-dependent SANS here to probe: the development of long-range FM order via low-$q$ scattering from FM domains and domain walls; the low-$T$ FM domain state via $q_x$-$q_y$ maps of the magnetic scattering after ZF-cooling; the nature of the paramagnetic to FM phase transitions via high-$q$ quasielastic scattering due to spin fluctuations; and potential phase separation into coexisting FM and paramagnetic regions. This is achieved through unpolarized SANS measurements from 3 to 60 K, in fields up to 6 T (parallel to the neutron beam), on representative $x$ = 0, 0.1, and 0.2 crystals. At $x$ = 0.1 and 0.2, the $c$-axis of the crystal was oriented both parallel and perpendicular to the neutron beam, to better assess the FM domain state.

Shown first in Figs.~\ref{fig:SANSmaps}(a,b) are low-$q$ range, $q_x$-$q_y$ ``maps'' of the low-$T$ (3 K) scattering from an $x$ = 0.1 crystal in both (a) zero and (b) 5-T field, taken after ZF cooling. As shown in Fig.~\ref{fig:SANSmaps}(a), these data were acquired with the crystal $c$-axis along $q_x$, resulting in a strong streak of scattering at an in-plane angle of $\sim$45$^{o}$. As detailed below, the scattering cross-section (d$\Sigma$/d$\Omega$) in this $q$ range decreases rapidly with $T$, and flattens above $T_C$, indicative of a substantial magnetic component. The data in Fig.~\ref{fig:SANSmaps} have thus had $T > T_C$ data subtracted to isolate the magnetic contribution. The magnetic origin of this scattering is further verified in Fig.~\ref{fig:SANSmaps}(b), where applying 5 T along the beam direction suppresses d$\Sigma$/d$\Omega$ to negligible levels, due to the absence of any FM magnetization perpendicular to the beam \cite{Sebastian2019}. The strong streak of scattering in Fig.~\ref{fig:SANSmaps}(a) is thus clearly of magnetic origin, due to the specific FM domain state after ZF-cooling, presumably dictated in large part by the magnetic anisotropy. At these $q$ values, the relevant FM domains and domain walls are on 30-300 nm length scales. Apparent from Fig.~\ref{fig:SANSmaps}(a) is thus that, at these length scales, significant scattering arises from FM magnetization that is not oriented along the $c$-axis, indicative of a possible weakening of the easy $c$-axis with $x$. For $x = 0.2$, Fig.~\ref{fig:SANSmaps}(c) reveals a different situation. Specifically, in order to generate significant $T$- and $H$-dependent scattering in this $q$ range, this crystal had to be oriented with the $c$-axis parallel to the beam. This then results in a broader streak of scattering at a different angle in the $q_x$-$q_y$ plane. Again, application of a 6 T magnetic field along the beam suppresses the scattering (Fig.~\ref{fig:SANSmaps}(d)) and hence confirms its magnetic origin. While full elucidation of the FM domain state would require extensive angle-dependent SANS measurements, ideally coupled with magnetic microscopy, it is clear from the comparison of Figs.~\ref{fig:SANSmaps}(a) and (c) that the domain state after ZF cooling is distinctly different at $x$ = 0.1 and 0.2. We thus clearly see $x$-dependent changes in the magnetic anisotropy and investigate this further with magnetometry in Section~\ref{section:easyaxis}. 

\begin{figure}
\includegraphics[width=0.45\textwidth]{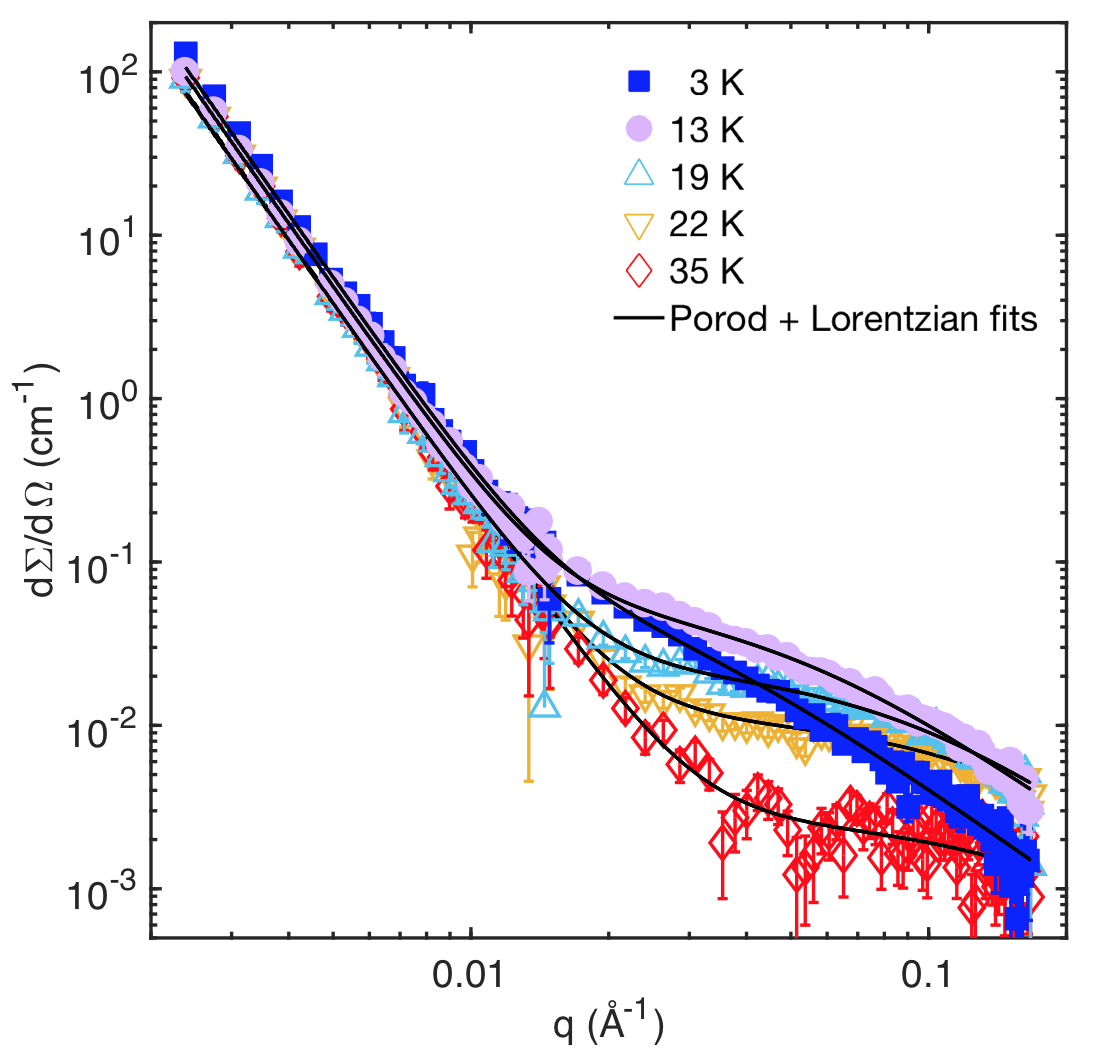}
\caption{Log$_{10}$-log$_{10}$ plot of the scattering wave-vector ($q$) dependence of the SANS cross section d$\Sigma$/d$\Omega$ for $x = 0.1$ at 3 K, 13 K (solid symbols below $T_C$), 19 K, 22 K and 35 K (open symbols above $T_C$). The solid lines are fits to a sum of Porod and Lorentzian functions, as described in the text. Note that the high-temperature data have not been subtracted here.}
\label{fig:SANSfits}
\end{figure}

\begin{figure}
\includegraphics[width=0.4\textwidth]{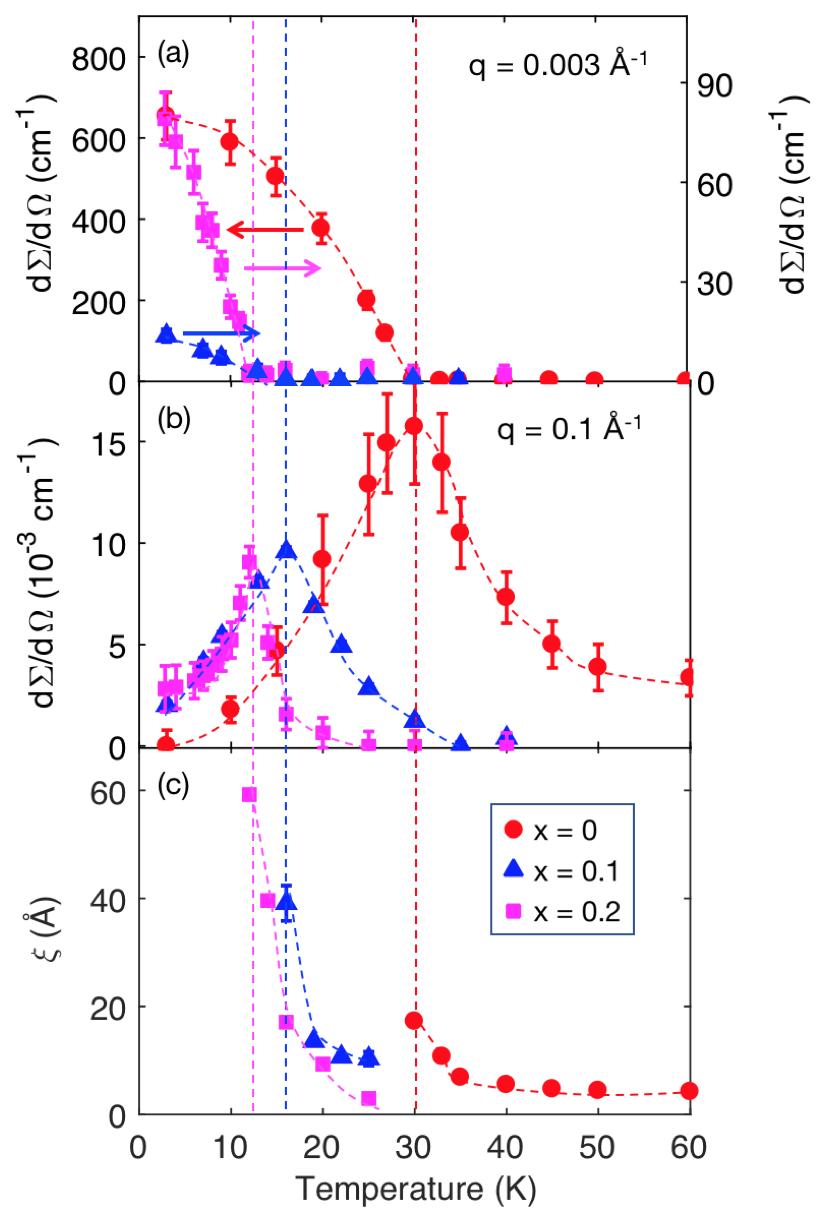}
\caption{(a,b) Temperature dependence of the SANS cross section (d$\Sigma$/d$\Omega$) for $x$ = 0, 0.1 and 0.2, at low $q$ (0.003 \AA$^{-1}$) and at high $q$ (0.1 \AA$^{-1}$), respectively. To improve statistics, the data at each $q$ are taken as averages of 6 adjacent $q$ points in the ranges 0.00238 \AA$^{-1}$ $<q<$ 0.00422 \AA$^{-1}$ and 0.09703 \AA$^{-1}$ $<q<$ 0.1106 \AA$^{-1}$, respectively. (c) $T$ dependence of the magnetic correlation length $\xi$ for $x$ = 0, 0.1 and 0.2 extracted from the Lorentzian term in Eq.~\eqref{eq:SANS}. Dashed lines are guides to the eye. Note that the high-temperature non-magnetic background has been subtracted in (a,b). However, this was not possible for the $x = 0$ data in (b) due to the non-zero Lorentzian scattering at 60 K.}
\label{fig:SANSparams}
\end{figure}

Figure~\ref{fig:SANSfits} shows (d$\Sigma$/d$\Omega$)($q$) profiles from an example $x$ = 0.1 crystal at 3, 13, 19, 22, and 35 K, i.e., representative temperatures from below to above $T_C$ ($\sim$ 16 K in this case). These data were obtained by radial averaging of $q_x$-$q_y$ maps (as in Figs.~\ref{fig:SANSmaps}(a,c)), a process that produces results similar to sector averaging around the magnetic scattering streak. Note, however, that no high-$T$ non-magnetic background has been subtracted in Fig.~\ref{fig:SANSfits}. It can be seen from the figure that from the lowest $q$ probed ($\sim$ 0.002 \AA$^{-1}$) to $\sim$ 0.01 \AA$^{-1}$, linear scaling occurs on this log$_{10}$-log$_{10}$ plot, i.e., we observe power-law behavior. The data can be fit to (d$\Sigma$/d$\Omega$)($q$,$T$) = (d$\Sigma$/d$\Omega$)$_P$($T$)/$q^n$, with $n$ = 4, and (d$\Sigma$/d$\Omega$)$_P$ being a $T$-dependent constant. This is the standard Porod-law form and the exponent $n$ = 4 indicates scattering from three-dimensional objects with smooth surfaces and size $\gg$ $2\pi/0.002$ \AA$^{-1}$ (the minimum $q$ probed), i.e., length scales larger than 300 nm \cite{Sebastian2019}. At low $T$ such as the 3 K data in Fig.~\ref{fig:SANSfits} (blue squares), these objects are domains/domain walls associated with the FM order \cite{Sebastian2019}, which is apparently truly long-ranged, consistent with our neutron diffraction and $\mu$SR results. As $T$ is increased, the scattering in this Porod regime decreases up to $T_C$ ($\sim$16 K), above which it becomes $T$-independent, as is typical, due to scattering from long-range structural inhomogeneities and defects \cite{Sebastian2019}. Distinctly different behavior occurs above $\sim$ 0.01 \AA$^{-1}$, where d$\Sigma$/d$\Omega$ adopts a weaker $q$ dependence, again strongly $T$-dependent. d$\Sigma$/d$\Omega$ in this $q$ regime first increases with increasing $T$, up to $T_C$, and then decreases to low levels at 35 K. The cross-section in this $q$ region can be fit to a Lorentzian form, i.e., (d$\Sigma$/d$\Omega$)($q$,$T$) = (d$\Sigma$/d$\Omega$)$_L$($T$)/[$q^2 + (1/\xi(T))^2$], where (d$\Sigma$/d$\Omega$)$_L$ is a $T$-dependent constant, and $\xi$ is the magnetic correlation length \cite{Sebastian2019}. This behavior is typical around FM to paramagnetic phase transitions, and occurs due to quasielastic scattering from short-range FM spin fluctuations \cite{Sebastian2019}. Such scattering typically grows on cooling from above $T_C$, then drops rapidly below $T_C$, which results in the ``critical scattering peak'' discussed below \cite{Sebastian2019}. Based on the above, the solid lines through the data in Fig.~\ref{fig:SANSfits} are fits to a sum of Porod and Lorentzian terms \cite{Bhatti2012,El-Khatib2019},

\begin{equation}
\frac{d\Sigma}{d\Omega}(q,T) = \frac{(\frac{d\Sigma}{d\Omega})_{P}(T)}{q^n} + \frac{(\frac{d\Sigma}{d\Omega})_{L}(T)}{q^2 + (1/\xi(T))^2},
\label{eq:SANS}
\end{equation}
with $n = 4$, which provides an excellent description of the data; this is also the case for $x$ = 0 and 0.2. 

Further analysis of the type of (d$\Sigma$/d$\Omega$)($q$,$T$) behavior shown in Fig.~\ref{fig:SANSfits} is provided in Fig.~\ref{fig:SANSparams}, which shows data for $x$ = 0, 0.1, and 0.2 crystals, with the high-$T$ non-magnetic scattering now subtracted. Shown first in Fig.~\ref{fig:SANSparams}(a) is (d$\Sigma$/d$\Omega$)($T$) at a fixed low $q$ $\approx$ 0.003 \AA$^{-1}$, where Porod scattering from long-range-ordered FM domains dominates. At $x$ = 0, the magnetic SANS cross-section turns on relatively sharply at $T_C$ $\sim$ 30 K, grows in an order-parameter-like fashion on cooling, consistent with the neutron diffraction results in Fig.~\ref{fig:NeutronAFM}, and then saturates at $\sim$700 cm$^{-1}$. The behavior for $x$ = 0.1 and 0.2, however, is distinctly different. The onset of magnetic scattering at the respective $T_C$ values remains apparent, but now with a quite linear increase in (d$\Sigma$/d$\Omega$) on cooling, and a low-$q$ cross-section reaching only $\sim$15-80 cm$^{-1}$. The order-parameter-like $T$ dependence is thus lost, and the absolute magnitude of the scattering from long-range FM domains is an order of magnitude lower than for $x$ = 0. Additional insight is gained from Fig.~\ref{fig:SANSparams}(b), which shows (d$\Sigma$/d$\Omega$)($T$) in the high-$q$ (Lorentzian-dominated) region. For $x$ = 0, the behavior appears relatively conventional \cite{Sebastian2019}, with a rather typical critical scattering peak around $T_C$, and a vanishing (d$\Sigma$/d$\Omega$) on cooling to 3 K. For $x$ = 0.1 and 0.2, however, the situation is again fundamentally different. Whereas clear peaks arise in both cases, it becomes apparent with increasing $x$ that the quasieleastic critical scattering does not vanish as $T \rightarrow 0$. At $x$ = 0.2, (d$\Sigma$/d$\Omega$)($q$ = 0.1 \AA$^{-1}$) in fact saturates at the lowest $T$. Fig.~\ref{fig:SANSparams}(c) further clarifies the situation by showing the extracted magnetic correlation length, $\xi$, which is seen to grow on cooling in each case, and to reach maximum values at $T_C$. Importantly, however, and contrary to expectations for a conventional second-order paramagnetic to FM phase transition \cite{Sebastian2019}, $\xi$ does not diverge, and instead exhibits a $T$ dependence on approach to $T_C$ that is notably weak. 
Drawing together the observations from Fig.~\ref{fig:SANSparams}, a picture emerges that is in excellent agreement with the neutron diffraction and $\mu$SR results. Specifically, while the FM order detected here by SANS is clearly long-ranged, even at $x$ = 0 the behavior is different from a completely conventional second-order paramagnetic to FM transition. The transition is in fact potentially weakly first-order, as inferred from the $\mu$SR data. Note that this weakly first-order nature of the transition becomes increasingly obvious with substitution $x$. Explicitly, critical fluctuations are certainly observed by SANS (Fig.~\ref{fig:SANSparams}(b,c)), but without the expected power-law divergence of $\xi$ as $T$ $\rightarrow$ $T_C$ \cite{Sebastian2019}. Such behavior persists at $x$ = 0.1 and 0.2, where, particularly at $x$ = 0.2, another notable feature emerges, namely the non-vanishing quasielastic critical scattering at even the lowest temperature (Fig.~\ref{fig:SANSparams}(b)). In light of the muon results implicating FM/paramagnetic phase coexistence at this $x$, we straightforwardly interpret this in terms of a small volume fraction of the crystal that remains in the paramagnetic phase even at the lowest temperature probed. These SANS measurements thus provide further evidence for the FM/paramagnetic phase coexistence at higher $x$. Finally, as is clear from comparing the three $x$ values in Fig.~\ref{fig:SANSparams}(a), our SANS data also provide clear support for the conclusion of substantially suppressed ordered FM magnetization with increasing $x$, which is seen to result in significantly suppressed Porod scattering from domains of long-range FM order. We note for completeness that the $T_C$ for the $x = 0.2$ crystal here is higher than those studied with magnetometry, neutron diffraction and $\mu$SR, likely due to oxygen off-stoichiometry.


\subsection{Easy-Axis Rotation}
\label{section:easyaxis}

\begin{figure*}
\includegraphics[width=0.95\textwidth]{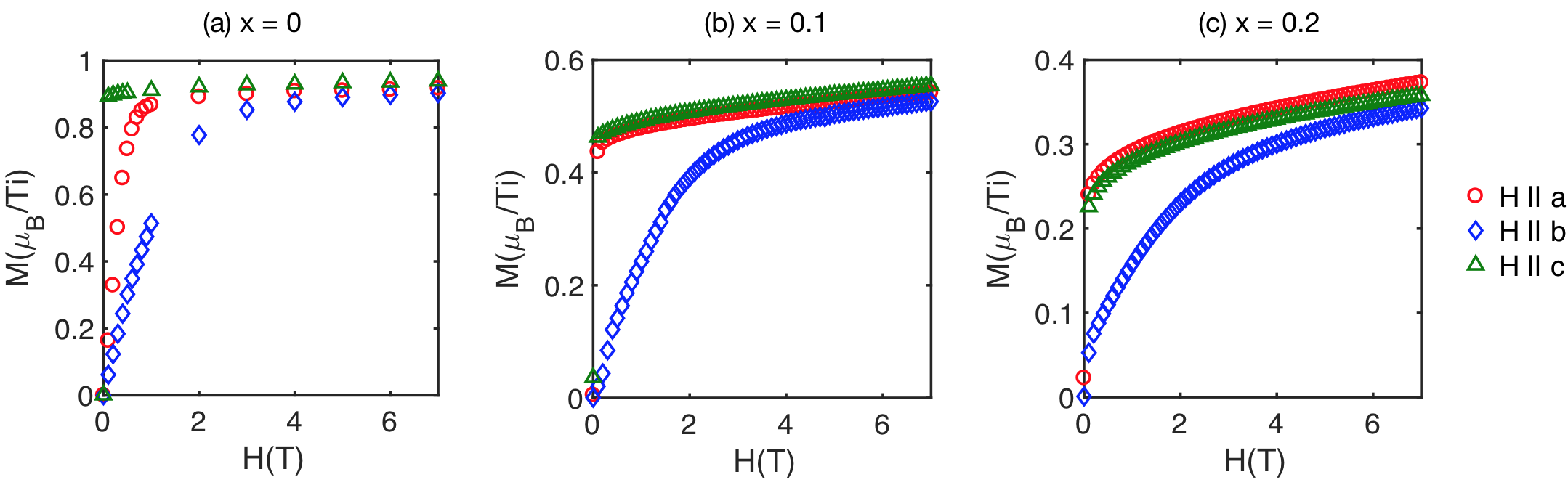}
\caption{Magnetic field dependence of the magnetic moment at 2 K for Y$_{1-x}$La$_{x}$TiO$_{3}$ with (a) $x = 0$, (b) $x = 0.1$ and (c) $x = 0.2$ along the three orthorhombic crystalline axes. The samples were zero-field cooled and measured in an ascending field.}
\label{fig:MagAnisotropy}
\end{figure*}

Given the significant differences in the FM domain state at $x$ = 0.1 and 0.2 observed by SANS in Fig.~\ref{fig:SANSmaps}, we investigated the changes in magnetic anisotropy with substitution using magnetometry. Figure~\ref{fig:MagAnisotropy} shows the magnetic moment as a function of magnetic field along the three orthorhombic crystalline axes for three different La-substitution levels, $x$ = 0, 0.1 and 0.2. The samples were cut from the same growths as those in Fig.~\ref{fig:Magn}, and polished into cubes with the $a$-, $b$- and $c$-axes perpendicular to one of the three inequivalent surfaces of the cube. 
The demagnetization factors for a regular cube-shaped sample are known \cite{Chen2002} and were taken into account. 
The magnetic moments observed here at 7 T along the $c$-axis are in agreement with those in Fig.~\ref{fig:Magn}. 
For $x = 0$, it is clear that the $c$- and $b$- axes are the easy and hard axes of magnetization, respectively, in agreement with previous reports based on floating-zone-grown YTiO$_{3}$ single crystals \cite{Tsubota2000,Kovaleva2007}. For $x = 0.1$ and 0.2, however, the magnetic anisotropy between the $c$- and $a$- axes is strongly suppressed compared to $x$ = 0, indicative of an easy plane. Similar to Fig.~\ref{fig:Magn}, the magnetic-field dependence of the magnetic moment is seen to significantly change with substitution, with YTiO$_{3}$ approaching complete saturation already at 7 T, in contrast to the situation for $x = 0.1$ and 0.2.

Although we clearly see a change in the magneto-crystalline anisotropy with La substitution, it must be noted that YTiO$_{3}$ single crystals grown with the Czochralski method show highly diminished magnetic anisotropy between the $a$- and $c$-axes, similar to the La-substituted single crystals grown here using the floating-zone method \cite{Garrett1981}. We also note that such strongly reduced anisotropy between the $a$- and $c$-axes is known to occur upon charge-carrier doping of YTiO$_{3}$ by substituting Ca at the Y site \cite{Tsubota2000}. In the literature, rotation of crystallographic easy axes in single crystals has been reported mostly as a function of temperature and magnetic field \cite{Zhou2005b,Kang2017} and attributed to intricate structural details as well as coupling between different magnetic sub-lattices. It is a distinct possibility that different methods of crystal growth cause subtle differences in single-ion anisotropy, which would in turn control the magneto-crystalline anisotropy. On the other hand, the observed changes in magnetic anisotropy could also be a consequence of strong spin-orbital coupling that is sensitive to the change in distortion of the TiO$_{6}$ octahedra as the average rare-earth ionic radius increases with La-substitution $x$ in Y$_{1-x}$La$_{x}$TiO$_{3}$ \cite{Goral1982}.

\begin{figure}
\includegraphics[width=0.45\textwidth]{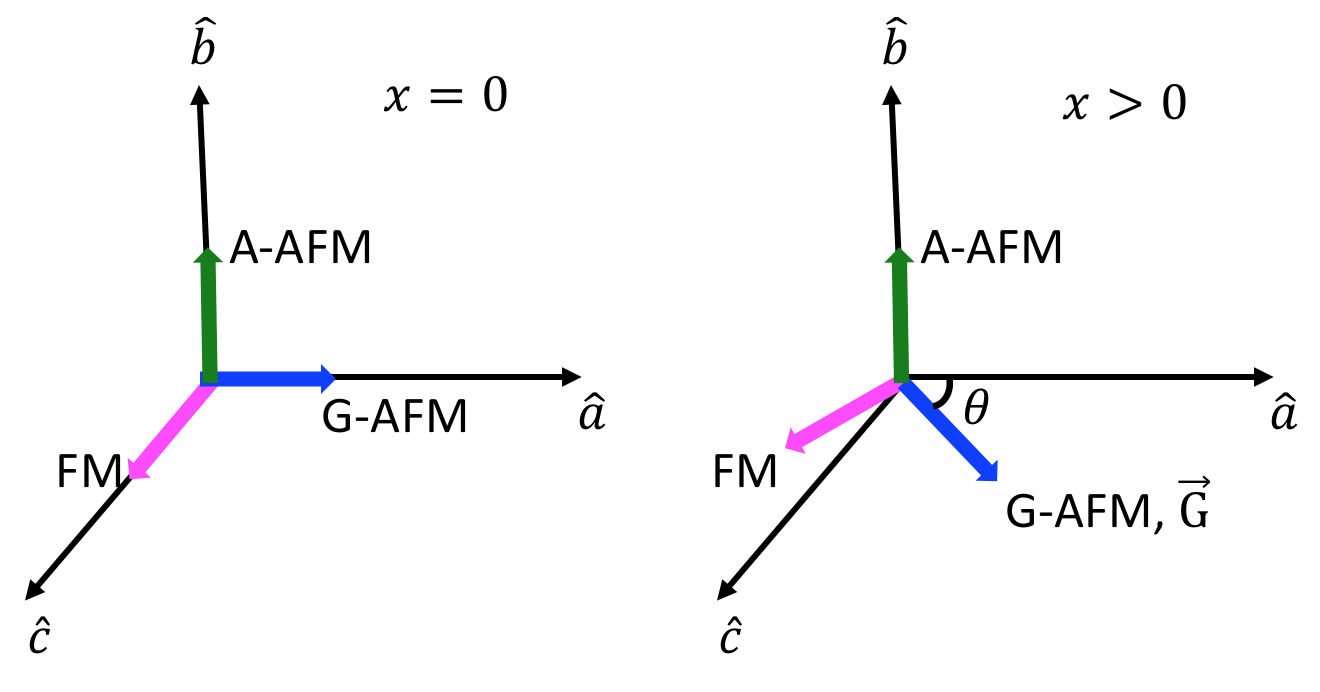}
\caption{Schematic of the ordered magnetic components in Y$_{1-x}$La$_{x}$TiO$_{3}$.}
\label{fig:Schematic}
\end{figure}

We now discuss the implications of the La substitution-induced changes observed in the magnetocrystalline anisotropy on the neutron diffraction results. In Section~\ref{section:neutron}, we deduced the substitution dependence of different magnetic components by comparing the intensity of specific magnetic Bragg reflections. If the magnetic structure remained invariant, this would be valid. However, given the changes in magnetocrystalline anisotropy, this may not be the case. Figure~\ref{fig:Schematic} shows a schematic of the changes in magnetic component directions upon La substitution. As noted in Section~\ref{section:neutron}, the FM, G-AFM and A-AFM moments were obtained from neutron intensities of (020)$_o$, (011)$_o$ and (001)$_o$ Bragg reflections, respectively. Crucially, only the spin component perpendicular to the Bragg reflection contributes to the measured magnetic intensity. In YTiO$_{3}$, the FM, G-AFM and A-AFM components point along $c$-, $a$- and $b$-axes, respectively, \cite{Ulrich2002} and therefore are perpendicular to the corresponding measured Bragg-reflections. Upon La substitution, changes are only observed in the magnetic anisotropy between $a$- and $c$-axes whereas the $b$-axis remains the hard-axis. Therefore, we assume that the A-AFM component continues to orient along the $b$-axis in the La-substituted system and hence the A-AFM Bragg reflection (001)$_o$ captures the entire A-AFM moment for all La-substitution levels. On the other hand, the FM spin component rotates in the $a$-$c$ plane upon substitution and no longer remains along the $c$-axis, as shown in Figure~\ref{fig:Schematic}. The FM Bragg reflection (020)$_o$, being directed along the $b$-axis, however, remains perpendicular to the FM spin component, and therefore still captures the entire FM moment even in the substituted system. Thus, the substitution dependence of the A-AFM and FM moments in Fig.~\ref{fig:FMmoments} remains unaffected by the observed changes in magnetocrystalline anisotropy. The (011)$_{\text{o}}$ Bragg reflection measures the G-AFM component of the ordered magnetic moment perpendicular to [011]$_{\text{o}}$. Considering that the G-AFM component, which was originally along the $a$-axis at $x = 0$, would now point along some direction in the $a$-$c$ plane (perpendicular to the FM component in the $a$-$c$ plane and the A-AFM component along the $b$-axis), the (011)$_{\text{o}}$ Bragg reflection will no longer capture the entire G-AFM moment for $x \neq 0$. We can nevertheless estimate the minimum fractional projection of the G-AFM magnetic moment component in the plane perpendicular to [011]$_{\text{o}}$.

As shown in Fig.~\ref{fig:Schematic}, let us assume that the G-AFM component $\vec{G}$  lies in the $a$-$c$ plane at an angle $\theta$ with respect to the $a$-axis. Let $a$, $b$ and $c$ be the lattice parameters of the orthorhombic crystalline lattice. Let $\hat{z}$ be the unit vector along [011]$_{\text{o}}$. The component of $\vec{G}$ perpendicular to [011]$_{\text{o}}$ can then be calculated as

\begin{equation}
G_{\perp} = \vec{G} - (\vec{G}\cdot\hat{z})\hat{z}.
\label{eq:4}
\end{equation}

Minimizing this equation, one can easily find that the component of $\vec{G}$ perpendicular to [011]$_{\text{o}}$ is smallest when $\theta = 90^{o}$, i.e., when the G-AFM component points along the $c$-axis. We can calculate this minimum component as

\begin{equation}
G_{\perp,\text{min}} = \lVert G \rVert \sqrt  \frac{1/b^2}{1/b^2+1/c^2}. 
\label{eq:5}
\end{equation}

Using the lattice parameters from Ref.\cite{Goral1982}, we obtain $G_{\perp,\text{min}}$ $\sim$ 0.8$\lVert G \rVert$ in the La-substitution range that we have studied. Figure~\ref{fig:neutronmoments} clearly shows that the measured fraction of G-AFM moment decreases to $< 0.8$ already for $x = 0.1$ and to $< 0.5$ for $x = 0.2$ and $0.25$, compared to $x = 0$, which implies that the conclusion regarding the G-AFM ordered moment being suppressed with substitution is still intact. However, the exact substitution dependence of the G-AFM ordered moment in Fig.~\ref{fig:neutronmoments} is questionable. The changes in the magnetic structure also contribute to the observed differences in the La-substitution dependence of the ordered moments measured by $\mu$SR compared to the bulk probes in Fig.~\ref{fig:LocalvsBulk}. 


\section{Conclusions}
\label{section:concl}

In conclusion, the magnetically-ordered ground state of single crystals of Y$_{1-x}$La$_{x}$TiO$_{3}$ for x $\leq$ 0.3 was studied using magnetometry, XAS/XMCD, 
neutron diffraction, $\mu$SR and SANS. 
We find that the bulk FM ordered moment and the canting-induced G-type and A-type AFM ordered moments are strongly suppressed on approaching the FM-AFM phase transition at the La-substitution $x_c$ $\sim$ 0.3. 
The suppression of the bulk ordered moment is due to a combination of reduced local ordered moment and phase separation into paramagnetic and magnetically-ordered regions, with a paramagnetic volume fraction of ~ 20\% at $x$ = 0.2 and 0.3. 
The absence of a clear signature of dynamic critical behavior for $x = 0 - 0.3$ and of a power-law divergence of the magnetic correlation length for  $x = 0 - 0.2$ suggests that the thermal phase transition is not conventional second-order. This becomes increasingly obvious with substitution $x$, as seen from both $\mu$SR and SANS measurements. In particular, for $x = 0.2$ and 0.3, $\mu$SR clearly observes a volume-wise separation into paramagnetic and magnetically-ordered regions near the respective transition temperatures, indicative of a first-order transition. Note, however, that our neutron diffraction data for $x = 0$ are consistent with critical behavior with the expected Ising critical exponent; the thermal phase transition for the unsubstituted system is therefore likely only very weakly first-order, as noted in Section~\ref{section:muSRvsNeutron}. An additional important aspect of La-substitution in Y$_{1-x}$La$_{x}$TiO$_{3}$ that we find here is the evidence for an evolution of the magneto-crystalline anisotropy from easy-axis to easy-plane. The results obtained here reveal complex changes in the magnetic structure upon approaching the FM-AFM phase boundary.

\section{Acknowledgments}
The work at University of Minnesota was funded by the Department of Energy through the University of Minnesota Center for Quantum Materials, under Grant No. DE-SC0016371. 
Parts of this work were carried out in the Characterization Facility, University of Minnesota, which receives partial support from NSF through the MRSEC program. 
Electron microprobe analyses of the crystal chemical composition were carried out at the Electron Microprobe Laboratory, Department of Earth Sciences, University of Minnesota-Twin Cities. 
A portion of this research used resources at the High Flux Isotope Reactor, a DOE Office of Science User Facility operated by the Oak Ridge National Laboratory. We acknowledge the support of the National Institute of Standards and Technology, U.S. Department of Commerce, in providing the neutron research facilities used in this work.
This research used resources of the Advanced Photon Source, a U.S. Department of Energy Office of Science User Facility operated for the DOE Office of Science by Argonne National Laboratory under Contract No. DE-AC02-06CH11357.
S. El-Khatib acknowledges travel support from AUS (Grant No. FRG17-T-06). 
Work at Columbia University was supported by US NSF DMR-1610633, the Reimei Project from the Japan Atomic Energy Agency, and Friends of U Tokyo Inc.

\bibliography{YLTOPaper.bib}
\end{document}